\documentclass[12pt]{article}

\usepackage{latexsym}

\usepackage[font=small,labelfont=bf]{caption}

\usepackage{graphics}
\usepackage{graphicx}
\usepackage{epstopdf}
\usepackage{amssymb}
\usepackage{caption}
\usepackage{subcaption}
\usepackage{tabularx}
\usepackage{gensymb}
\usepackage{mathtools,amsmath,amssymb,mathrsfs,color,esint}
\usepackage{array}
\usepackage{makecell}
\usepackage{float}
\usepackage{dcolumn}
\usepackage{hyperref}

\begin{document}

\title{Polarized equatorial emission and hot spots around black holes with a dark matter halo}

\author{
Tsanimir Angelov$^{1}$\footnote{E-mail: \texttt{tsangelov@phys.uni-sofia.bg}}, \,
Rasim  Bekir$^{1}$\footnote{E-mail: \texttt{rbekir@uni-sofia.bg}}, \,
Galin Gyulchev$^{1}$\footnote{E-mail: \texttt{gyulchev@phys.uni-sofia.bg}},  \, \\
Petya Nedkova$^{1}$\footnote{E-mail: \texttt{pnedkova@phys.uni-sofia.bg}}, \,
Stoytcho Yazadjiev$^{1,2}$\footnote{E-mail: \texttt{yazad@phys.uni-sofia.bg}}\\ \\
 {\footnotesize${}^{1}$ Faculty of Physics, Sofia University ``St. Kliment Ohridski'',}\\
  {\footnotesize 5 James Bourchier Boulevard, Sofia~1164, Bulgaria } \\
    {\footnotesize${}^{2}$ Institute of Mathematics and Informatics,}\\
{\footnotesize Bulgarian Academy of Sciences, Academic G. Bonchev 8, } {\footnotesize Sofia 1113, Bulgaria}}
\date{}
\maketitle

\begin{abstract}
We study the linear polarization of the accretion disk around black holes with a dark matter halo. The interaction of the black hole with the dark matter is modelled by considering an exact solution to the Einstein equations which describes a superposition of the Schwarzschild black hole with a Hernquist-type matter distribution. We simulate the observable polarization of a magnetized fluid ring orbiting around the black hole and evaluate the influence of the dark matter halo on its properties for physical parameters compatible with the dark matter distribution in galaxies. The polarization intensity and direction of the direct images deviate with less than $1\%$ from the isolated Schwarzschild black hole for a range of magnetic field configurations. For the indirect images the deviation increases with an order of magnitude but still remains under $10\%$ for small inclination angles corresponding to the galactic targets M87* and SgrA*. This makes the detection of the dark matter impact on the polarized emission from the accretion disk extremely challenging in the near future.
\end{abstract}

\section{Introduction}

Recent advances in the experiments in the electromagnetic spectrum opened new ways of testing gravity in the strong field regime. The observational data from the Event Horizon Telescope (EHT) and the interferometer GRAVITY provided access to a range of properties of the gravitational field around the supermassive compact objects at the center of galaxies which may be used to get insights about their nature \cite{EHT1}-\cite{GRAVITY_2}. These include the shadow of the compact objects and the emission of the accretion disk surrounding them which gives rise to observational images with different morphology and polarization.

An important fundamental question is to uncover the nature of astrophysical compact objects as inferred by the available observational data. They may be described by the Kerr solution in general relativity but they may also be manifestations of black holes in some modified theories of gravity. Furthermore, we may observe horizonless compact objects carrying hints about an effective theory of quantum gravity. These scenarios are encoded in the electromagnetic observables with different sensitivity which may bring difficulties to distinguish among them in some of the observational channels.

Another issue which arises when interpreting observational data is whether we describe adequately the gravitational interaction of the compact objects with its environment. The experimental data from the EHT and GRAVITY missions is usually modelled by an accretion flow propagating in the gravitational field of the compact object at the galactic center. In this construction the back reaction of the accreting matter on the spacetime geometry is neglected as well as the influence of galactic dark matter halo. These effects are typically considered small but they may become significant if high precision experimental data is available. In particular, they may be comparable with effects connected with the non-Kerr nature of the central compact object and create bias in the interpretation of the observational data.

In this paper we investigate the influence of the galactic dark matter halo on the linear polarization of the emission from the accretion disk. For the purpose we consider a static spherically symmetric solution to the Einstein equations with an anisotropic matter distribution which contains a black hole at small scales and a halo with a Hernquist-type density distribution at large scales \cite{Hernquist:distribution}, \cite{Cardoso}. Similar density profiles are well-known to be consistent with the dark matter simulations in galaxies and to reproduce the galactic rotation curves. The Schwarzschild solution with a dark matter halo was previously applied for studying the dark matter impact on other phenomenological effects such as the propagation of gravitational waves \cite{Cardoso}, the black hole shadow \cite{Xavier:2023}, or the observational appearance of thin accretion disks \cite{Macedo:2024}. We continue this line of research by investigating the polarization of the equatorial emission in such spacetimes and accessing the deviation in its observational properties produced by the dark matter content.

For the purpose we describe the polarized emission from the inner regions of the accretion disk by considering an analytical toy model of a thin fluid ring orbiting in the equatorial plane around the central compact object proposed in \cite{Narayan:2021}, \cite{Gelles:2021}. The fluid is located in a constant magnetic field and emits synchrotron radiation. Despite its simplicity this model manages to reproduce the observed polarization from the inner accretion disk for the galactic target M87* with a reasonable precision. It can be further applied for describing the polarization properties of the emission of the hot spots in the vicinity of the compact object at the center of our galaxy Sgr A*. The hot spots represent highly localized flares with approximately ten times higher intensity than the quiescent flux of the accretion disk of SgrA* which propagate on circular orbits located at 8-10 gravitational radii. In the case of the near-infrared flares measured by GRAVITY the background emission is insignificant and the polarized flux from the hot-spot can be reasonably described by the model of synchrotron emitting ring. The polarization properties of various spacetimes were investigated previously within this framework including black holes in the modified theories of gravity \cite{Qin:2021}-\cite{Zhang:2022} as well as horizonless compact objects \cite{Nedkova:2023}-\cite{Delijski}.

The paper is organized as follows. In the next section we present the solution of the Einstein equations coupled to anisotropic fluid which describes the Schwarzschild black hole surrounded by a dark matter halo. In section 3 we review the model of synchrotron emitting fluid orbiting in the equatorial plane. We further derive the observable properties of the polarized radiation and describe the procedure for their computation. In section 4 we present our simulations for the observable linear polarization for the Schwarzschild black hole with a dark matter halo and discuss the deviation in the polarization properties caused by the impact of the dark matter. We consider several magnetic field configurations justified by the observational data for M87* and Sgr A*. In the last section we summarize our results.

\section{Black holes surrounded by a dark matter halo}

We consider a spacetime geometry which describes a superposition of a black hole and a dark matter halo located at a large distance. The solution is inspired by the construction of the Einstein cluster, which was described first by Einstein in \cite{Einstein:Clusters1} and revisited in \cite{Einstein:Clusters2}. It represents a stationary self-gravitating system consisting of many massive particles which move on circular geodesics in the gravitational field created by their superposition. Such system possesses a spherical symmetry and  the metric can be represented in the form

\begin{equation}
ds^2=-e{^\nu}dt^2+ e{^\lambda}dr^2+r{^2}(d\theta^2 + \sin \theta d\phi^2)\,,\label{eq-lineelement}
\end{equation}
where the functions $\nu$ and $\lambda$ depend only on the radial coordinate. The Einstein construction assumes $n$ particles with mass $m{_p}$, 4-velocity $U{^\mu}$ and four momentum $P{^\mu} = m{_p}U^{\mu}$ that satisfy the geodesic equations. Then, we can introduce an averaged stress-energy tensor of all the particles

\begin{equation}
\langle T{^\mu}{^\nu}\rangle = \frac{n}{m{_p}}\langle P{^\mu}P{^\nu}\rangle,
\end{equation}
where we average over all the geodesic trajectories passing through the point where the stress-energy tensor is evaluated. Calculating the averaged stress-energy tensor with the described symmetries we obtain that the only  non-vanishing components are

\begin{align}
\langle T^t_t \rangle = -\rho , \qquad \langle T^\theta_\theta \rangle = \langle T^\phi_\phi\rangle = p,
\end{align}
where $\rho$ is the energy density of the system and $p$ is the tangential pressure. The corresponding Einstein equations possess the form
\begin{eqnarray}
&&\frac{1}{r^2}\left[r(1-e{^{- \lambda}})\right]{^{'}} = 8 \pi \rho\,, \\
&&\nu^{'} = \frac{1}{r}(e{^{- \lambda}}-1)\,, \\
&&\frac{e^{- \lambda}}{2}\left[\nu^{''}+\frac{\nu^{'2}}{2}+ \frac{\nu^{'}-\lambda^{'}}{r} - \frac{\nu^{'}\lambda{^{'}}}{2}\right] =  8 \pi p.\,
\end{eqnarray}
leading to the following relation between the energy density and the tangential pressure
\begin{eqnarray}\label{relation_pr}
p = \frac{\rho}{4}(e^\lambda-1) =\frac{r\nu^{'}}{4}\rho.
\end{eqnarray}

We can introduce the gravitational mass $m(r)$ of the system contained in a spherical region with radius $r$
\begin{equation}
m(r) = 4 \pi \int_{0}^{r} \rho r^2 \,dr,
\end{equation}
and express all the characteristics of the system in terms of this quantity. Using Einstein equations we can obtain its relation
to the metric function $\nu$
\begin{equation}
\frac{r}{2}\nu^{'} = \frac{m(r)}{r-2m(r)},
\end{equation}
and applying Eq. $\ref{relation_pr}$ we can further express the metric function $\lambda$ and the tangential pressure as
\begin{eqnarray}\label{Einstein_eq}
e^\lambda = (1-2m(r)/r)^{-1}, \quad~~~ p = \frac{m(r)}{2(r-2m(r))}\rho.
\end{eqnarray}

Due to their anisotropic energy-momentum tensor the Einstein clusters are suitable systems for  modeling the dark matter halos in galaxies. Any galactic rotation curve can be reproduced by selecting an appropriate tangential pressure profile, or alternatively by specifying the mass distribution $m(r)$ \cite{Lake}-\cite{Bomer:2007}. Thus, Einstein clusters have been constructed with King-type, Burkert-type and Navarro-Frenk-White  density distributions which provide a good approximation to the observed dark matter distributions in galaxies.

We can further specify a mass distribution which describes not only a dark matter halo but includes the interaction with the central galactic black hole. This idea was explored in \cite{Cardoso} where the following mass function was proposed

\begin{equation}\label{m(r)}
m(r)=M{_{\rm BH}}+\frac{Mr^2}{(a_{0}+r){^2}}\left(1-\frac{2M{_{\rm BH}}}{r}\right)^2\,,
\end{equation}
where $M$, $a_0$ and $M{_{\rm BH}}$ are constants. Integrating eq. $(\ref{Einstein_eq})$ we obtain
\begin{eqnarray}\label{eq-fradial}
&&e^{\lambda} = \left(1-\frac{2M{_{\rm BH}}}{r}\right)e^\Upsilon \,, \nonumber \\[2mm]
&&\Upsilon=-\pi\sqrt{\frac{M}{\xi}}+2\sqrt{\frac{M}{\xi}}\arctan\left({\frac{r+a{_0}-M}{\sqrt{M\xi}}}\right)\,, \nonumber \\[2mm]
&&\xi=2a{_0}-M+4M{_{\rm BH}}\,,
\end{eqnarray}
and we can express the matter density as

\begin{equation}
4\pi\rho=\frac{m'}{r^{2}}=\frac{2M(a{_0}+2M{_{\rm BH}})(1-2M{_{\rm BH}}/r)}{r(r+a{_0})^3}\,.\label{eq-hernquist_GR}
\end{equation}

The spacetime contains a horizon at $r=2M_{BH}$ and a curvature singularity at $r=0$. When the solution parameters obey  $M{_{\rm BH}} \ll M \ll a{_0}$ no other curvature singularities occur. Then the solution reduces to the Schwarzschild black hole with mass $M_{BH}$ at small radii and a Hernquist-type matter distribution \cite{Hernquist:distribution} at large radii and  can be interpreted as a black hole interacting with a dark matter halo. The parameters $M$ and $a_0$ correspond to the halo mass and its characteristic length scale while the ADM mass of the configuration is $M{_{\rm ADM}} = M + M{_{\rm BH}}$. Astrophysical settings are described adequately by the scale ordering $M{_{\rm BH}} \ll M \ll a{_0}$ as implied by cosmological probes of the dark matter content in galaxies.

The influence of the dark matter modifies the location of the photon sphere $r_{ph}$ and the ISCO $r_{ISCO}$ with respect to the Schwarzschild black hole. Approximate expressions up to terms $\sim {\mathcal{O}}(\frac{1}{a_0})$ were derived as \cite{Cardoso}
\begin{eqnarray}\label{rISCO_GDM}
r_{ph}\approx 3M_{\rm BH}\left(1+\frac{MM_{\rm BH}}{a_0^2}\right)\,, \nonumber \\[2mm]
r_{ISCO}\sim 6M_{\rm BH}\left(1-\frac{32MM_{\rm BH}}{a_0^2}\right)\,.
\end{eqnarray}
Then we can calculate the critical impact parameter on the null geodesics which determines the black hole shadow as

\begin{equation}
b_{\rm crit}=3\sqrt{3}M_{\rm BH}\left(1+\frac{M}{a_0}+\frac{M(5M-18M_{\rm BH})}{6a_0^2}\right)\,.
\end{equation}

Although the dark matter halo is characterized by two parameters its impact on the black hole environment and certain strong gravity phenomenology may be governed mostly by their ratio. It determines the compactness of the halo given by
\begin{equation}
C = \frac{M}{a{_0}}\,.
\end{equation}
In our study we are interested in the gravitational field of supermassive black holes in the center of galaxies which implies values of the compactness parameter in the range $10^{-3}<C<10^{-8}$ \cite{NFW}-\cite{Wang}.

\section{Polarized image of an equatorial emitting ring}

This work aims at describing the light reaching an observer from a thin fluid ring in the vicinity of a black hole with a dark matter halo. The fluid emits linearly polarized synchrotron radiation as it orbits at a fixed radial coordinate with a locally constant magnetic field permeating it. Parallel-transport of the polarization vector along the null geodesics is used to determine the observable polarization at large distances.

This technique has been used previously for the Schwarzschild and Kerr black holes in \cite{{Narayan:2021}, {Gelles:2021}}. The general static spherically symmetric case is described in \cite{Nedkova:2023}. Only a short summary adjusted to the specific metric of a black hole with a dark matter halo will be described in this paper. Considering the symmetries of the spacetime we define the orthonormal tetrad $\{e^{\mu}_{(a)}\}$ which will be used 

\begin{eqnarray}\label{eq:tetrad}
     &&e_{(t)} =\left(1-\frac{2 M_{\mathrm{BH}}}{r}\right)^{-1/2} e^{-\Upsilon/2}, \,\,\,\,\,
    e_{(r)} = \left(1-\frac{2m(r)}{ r}\right)^{1/2}\partial_r, \,\,\,\,\,
    e_{(\theta)} =  \frac{1}{r}, \nonumber \\[2mm]
    &&e_{(\phi)} = \frac{1}{r\sin\theta},
\end{eqnarray}
where equations \eqref{m(r)} and \eqref{eq-fradial}  are substituted.

The fluid moves only in the equatorial $(\hat{r})$-$(\hat{\phi})$ plane. Its velocity in the local frame has magnitude $\beta$ and direction $\chi$. In vector notation it can be represented as

\begin{equation} \label{eq:betaboost}
\vec{\beta} = \beta\left(\cos\chi\,(\hat{r})+\sin\chi\,(\hat{\phi})\right).
\end{equation}
The synchrotron emission can be described in the rest frame of the fluid i.e. the boosted emitter so we need to use a Lorentz transformation matrix to switch between frames

\begin{eqnarray}\label{Lorenz_tr}
\hat{e}^{\,\mu}_{(a)} = \Lambda^{\hspace{0.3cm}(b)}_{(a)}e^{\mu}_{(b)},
\end{eqnarray}

\begin{eqnarray}
    \small
    &&\Lambda
        =\begin{pmatrix}
             \gamma  & -\beta  \gamma  \cos \chi & -\beta  \gamma  \sin \chi & 0 \\
             -\beta  \gamma  \cos \chi & (\gamma -1) \cos ^2\chi+1 & (\gamma -1) \sin \chi \cos \chi & 0 \\
             -\beta  \gamma  \sin \chi & (\gamma -1) \sin \chi \cos \chi & (\gamma -1) \sin ^2\chi+1 & 0 \\
             0 & 0 & 0 & 1 \\
        \end{pmatrix}, \\[2mm]
    \normalsize
    &&\gamma =(1-\beta^2)^{-1/2}.
\end{eqnarray}
The magnetic field which permeates the fluid is given by $\vec{B} =\left(\hat{B}^{ r}, \hat{B}^{\phi}, \hat{B}^{\theta}\right)$. The polarization of the synchrotron emission is described by using the local 3-momentum $\vec{p} =\left(\hat{p}^{ r}, \hat{p}^{\phi}, \hat{p}^{\theta}\right)$ as

\begin{equation} \label{eq:polcross}
  \vec{f} = \frac{\vec{p} \times \vec{B}}{|\vec{p}|},
\end{equation}
where the cross product gives us three spacial components while $f^{(t)}=0$. We define the angle between the magnetic field and the fluid's 3-momentum as $\zeta$ where

\begin{equation} \label{zeta_B}
\sin \zeta = \frac{| \vec{p} \times \vec{B}|}{|\vec{p}||\vec{B}|}.
\end{equation}
Then we obtain the following normalization
\begin{equation}
\hat{f}^{a}\hat{f}_{a} = \sin^2\zeta |\vec{B}|^2.
\end{equation}
The transformation of the polarization 4-vector $f^{\mu}$ from the fluid rest frame to the rest frame of the black hole is done by using the Lorentz transformation ($\ref{Lorenz_tr}$).

In order to describe the observable polarization at large distances we need to parallel transport the polarization 4-vector along the null geodesics. Since $f^{\mu}$ remains normal to $p^{\mu}$ then it must satisfy the following equations
\begin{eqnarray}
&&p^\mu \nabla_{\mu} p_\nu =0 \nonumber \\[2mm]
&&p^{\mu}\nabla_{\mu}f_\nu=0, \;\;\; p^\mu f_\mu=0.
\end{eqnarray}
These differential equations can be solved algebraically given the case of a static spherically symmetric spacetime. This is due to the existence of hidden symmetries of the spacetime described by  irreducable Killing and Killing-Yano tensors \cite{Nedkova:2023}. First we will take advantage of the second order Killing-Yano tensor $Y_{\mu\nu}$. In the spacetime of black holes with dark matter halos it has the following non-zero components
\begin{eqnarray}
Y_{\theta\phi}=-Y_{\phi\theta}= r^3\sin\theta.
\end{eqnarray}
Another symmetry corresponds to the conformal Killing-Yano tensor $\tilde{Y}_{\mu\nu}$ generated by the Hodge duality $\tilde{Y}_{\mu\nu}=\frac{1}{2}\epsilon_{\mu\nu\alpha\beta}Y^{\alpha\beta}$. The constants of motion along null geodesics associated with these symmetries are 
\begin{eqnarray}
\kappa_1= \frac{1}{2} {\tilde Y}_{\mu\nu}p^\mu f^\nu , \;\;\;    \kappa_2= Y_{\mu\nu} p^\mu f^\nu.
\end{eqnarray}
In the spacetime of black hole with a dark matter halo they can be expressed by the polarization vector $f^{\mu}$ and the 4-momentum of the photon $p^{\mu}$ as 

\begin{eqnarray}\label{kappa_12}
&&\kappa_1= \left(\frac{1-2M_{BH}}{1-2m}e^{\Upsilon}\right)^{\frac{1}{2}}r (p^t f^r - p^rf^t), \nonumber  \\[2mm]
&&\kappa_2= r^3 \sin \theta (p^\theta f^\phi - p^\phi f^\theta).
\end{eqnarray}

Now that we have a complete algebraic representation of the polarization 4-vector as the photon travels throughout the spacetime we can calculate the observable polarization for a distant observer. First we need to define coordinates on their celestial sphere. The tetrad that was defined earlier allows us to define two angles on the observer's sky using the projection of the 4-momentum \cite{Bardeen}

\begin{eqnarray}
\sin\tilde\alpha = \frac{p^{(\theta)}}{p^{(t)}}, \quad \,\,\, \tan\tilde\beta = \frac{p^{(\phi)}}{p^{(r)}},
\end{eqnarray}
Finally we rescale the coordinates from $\tilde\alpha$ and $\tilde\beta$ to $x=r\tilde\alpha$ and $y=r\tilde\beta$ and obtain the celestial coordinates
\begin{eqnarray}\label{obs_coord}
x &=& -r\,p^{(\phi)} = -\frac{p_{\phi}}{\sin\theta_0}, \nonumber \\[2mm]
y &=& r\,p^{(\theta)} = p_{\theta},
\end{eqnarray}
where $\theta_0$ is the inclination angle of the observer.

We can obtain the projection of the polarization vector on the observer's frame in terms of the constants of motion
\begin{eqnarray}
f^x = \frac{x\kappa_1  + y\kappa_2}{x^2 + y^2} , \nonumber \\[2mm]
f^y = \frac{y\kappa_1 - x\kappa_2}{x^2 + y^2} ,
\end{eqnarray}
and normalize these components to unity
\begin{eqnarray}\label{obs_polarization}
f^x_{obs} = \frac{x\kappa_1  + y\kappa_2}{\sqrt{(\kappa_1^2 + \kappa_2^2)(x^2 + y^2)}} , \nonumber \\[2mm]
f^y_{obs} = \frac{y\kappa_1  - x\kappa_2}{\sqrt{(\kappa_1^2 + \kappa_2^2)(x^2 + y^2)}} .
\end{eqnarray}
Now that the polarization vector is expressed solely through the constants of motion $\kappa_1$ and $\kappa_2$ we can trace back their origin to the emission of the photon from the orbiting fluid in the disk around the central black hole. That would allow us to express the two constants using only the physical quantities present at the photon's emission point. We project the polarization vector as initially defined in the fluid rest frame \eqref{eq:polcross} in the local tetrad \eqref{eq:tetrad} and substitute $\kappa_1$ and $\kappa_2$. This leads us to the following expressions used in our simulations
\begin{eqnarray}\label{kappa_fr}
\kappa_1= \gamma r \left[{\hat p^{(t)}}{\hat f^{(r)}}  + \beta \left({\hat p^{(r)}}{\hat f^{(\theta)}} - {\hat f^{(r)}}{\hat p^{(\theta)}} \right) \right], \nonumber \\[2mm]
\kappa_2 = \gamma r \left[\left({\hat p^{(\theta)}}{\hat f^{(\phi)}} - {\hat f^{(\theta)}}{\hat p^{(\phi)}} \right)  - \beta {\hat p^{(t)}}{\hat f^{(\phi)}}\right],
\end{eqnarray}

We are interested in two observables in particular which represent the direction of the polarization vector on the observer’s sky and the intensity of the polarization. 

The polarization intensity is proportional to the norm of the observable polarization vector but further modifications are needed \cite{Narayan:2021}. They come from the fact that the synchrotron emission intensity is dependent on the angle $\zeta$ between the fluid's 3-momentu and the magnetic field, the photon's frequency, and the geodesic path length in the fluid
\begin{equation}\label{LenghtPath}
 l_p=\frac{\hat p^{(t)}}{\hat p^{(\theta)}}\,H,
\end{equation}
where $H$ is the disk thickness and $\hat{p}^{(\mu)}$ is the 4-momentum measured in the rest frame of the fluid. While the photon travels from the emission point to the observer it accumulates a Doppler boost which needs to be taken into account. Its effect is a modulation by a factor of $\delta^{3+\alpha_\nu}$, where by $\delta$ we denote the ratio between the energy of the photon in the observer's frame and the energy of the photon measured in the fluid frame  
\begin{equation}
\delta = \frac{E_{obs}}{E_s} = \frac{1}{\hat p^{(t)}},
\end{equation}
The obtained expression for the polarization intensity is \cite{Narayan:2021}

\begin{equation}
|I| = \delta^{3+\alpha_\nu}\, l_p \, (\sin\zeta)^{1+\alpha_\nu},
\end{equation}
where the spectral index will be chosen as $\alpha_\nu=1$ to be consistent with the observations of M87*  \cite{{Narayan:2021}, {Gelles:2021}}. Thus we obtain the components of the observed electric field 
\begin{eqnarray}\label{pol_EVPA}
E^x_{obs} &=& \delta^2\, l_p^{1/2}\, \sin\zeta\, f^x_{obs}, \nonumber \\[2mm]
E^y_{obs} &=& \delta^2\, l_p^{1/2}\, \sin\zeta\, f^y_{obs}, \nonumber  \\[3mm]
|I| &=& (E^x_{obs})^2 + (E^y_{obs})^2.
\end{eqnarray}

The direction of the polarization vector is determined by the electric vector position angle (EVPA) on the observer's sky 
\begin{equation}
    EVPA = \arctan\left({-\frac{f^x_{obs}}{f^y_{obs}}}\right).    
\end{equation}
We will measure the EVPA in positive (counter-clockwise) direction from the $x>0$ semi-axis.

\section{Linear polarization for black holes with a dark matter halo}

In this section, we analyze the linear polarization of synchrotron emission in the static spacetime of a black hole surrounded by a dark matter (DM) halo. The goal is to assess how dark matter influences the observable polarization pattern compared to the Schwarzschild case. We examine both the intensity and orientation of the polarization vector across different magnetic field configurations, with emphasis on the effect of halo compactness. 

Although the solution describing  a black hole with a dark matter halo depends on two parameters we observed in our simulations that the polarization properties are determined mainly by the halo compactness. We computed the polarization intensity and direction for several sets of configurations with the same compactness parameter in the range $C \in (10^{-8}, 10^{-3})$ but with different values of the length scale $a_0$ and the ratio $M/M_{BH}$ taking the ranges $a_0\in (10^5,10^{12})$ and $M/M_{BH}\in (10^2, 10^5)$. We observed that the deviation in the polarization intensity and direction within each sample with constant compactness is $\sim 10^{-5}$ or lower which is far below the sensitivity of near-future observational capacities. Therefore, the investigation of the polarization properties only with respect to the halo compactness brings little bias for the selected range of parameters. 

In all the simulations the observer is located at the radial coordinate $r= 4.10^{12}\, M_{BH}$. The emission radius of the synchrotron radiation for the Schwarzschild black hole is $r=4M$. In order to be able to make a comparison, the emission radius for the black holes with a dark matter halo is shifted accordingly so that they produce an observable image in the same location on the observer's sky as the emitting ring at $r=4M$ in the Schwarzschild spacetime. For all the selected values of the compactness parameter the shadow size of the black holes with a dark matter halo is consistent with the EHT constrains for the galactic targets M87* and Sgr A*.

\subsection{Direct images}

\begin{figure}[t!]
    \centering
    \includegraphics[width=14.2cm]{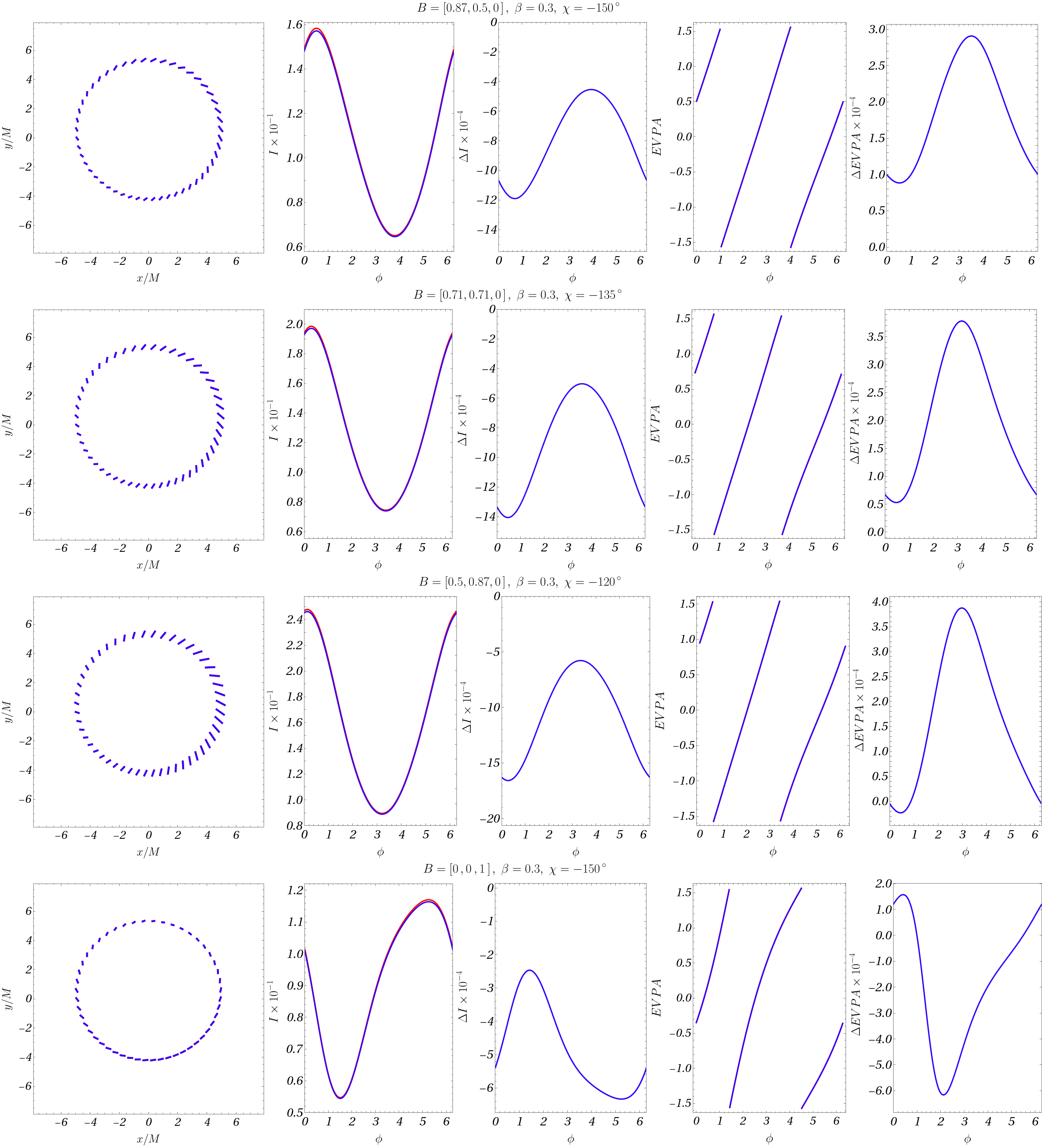}
    \caption{Linear polarization for a black hole with a dark matter halo with compactness parameter $C=10^{-3}$ (blue line) compared to the Schwarzschild black hole (red line). The inclination angle is $i = 20^\circ$ and the emission radius for the Schwarzschild spacetime is $r=4M$. We analyze the polarization intensity I and direction EVPA, and their deviation from the Schwarzschild black hole $\Delta \text{I}$ and $\Delta \text{EVPA}$ (see main text).}
    \label{fig:direct-inclination20}
\end{figure}

In this study, we analyze the polarization properties of synchrotron emission in spacetimes describing a black hole surrounded by a dark matter halo, under different magnetic field configurations. We focus on direct images, which are formed by photon trajectories that reach the observer without completing full loops around the compact object, and are thus characterized by azimuthal angles in the range $\phi \in [0, \pi)$. We begin by considering the case of a purely equatorial magnetic field, which shows better agreement with polarization observations of sources such as M87*. Simulations reveal that even in the presence of a dark matter halo, the polarization vector maintains its characteristic rotation angle, and the intensity distribution remains largely similar to that of a Schwarzschild black hole, with a peak on the right side of the image and a gradual decrease toward the poles. This behavior is primarily caused by gravitational redshift effects, which dominate at the edges of the image, particularly on the left and right, where the photon energy is most strongly affected by the gravitational potential (see Figure~\ref{fig:direct-inclination20}, rows 1–3, first column). Although the total intensity of the polarized flux decreases with increasing compactness $M/a_{0}$ of the dark matter distribution in the galactic halo (i.e., larger deviations from the Schwarzschild solution), the symmetry of the image is preserved and closely follows the case without dark matter.

In contrast, the vertical magnetic field configuration produces a noticeably different polarization pattern, as illustrated in the bottom row of Figure \ref{fig:direct-inclination20}. It is well established that vertical fields in the Schwarzschild geometry lead to polarization structures that are inconsistent with the observational data for M87*. This inconsistency arises from the combined effects of aberration, which is present in any static spacetime, and gravitational lensing, which depends on the specific geometry. Simulations involving black holes with a surrounding dark matter halo and vertical magnetic fields show that while the overall image structure and symmetry remain similar to the Schwarzschild case, the resulting polarization is again incompatible with the observations of astrophysical systems such as M87*. As the halo compactness increases, the polarization signal becomes significantly weaker, although the basic structure of the image, characterized by a maximum intensity at the bottom of the ring and a gradual decrease upwards, remains consistent with the Schwarzschild case. 

\begin{table}[h!]
    \centering
      \begin{tabular}{|c|c|c|c|c|}
       \hline
       \thead{ $i = 20^{\circ}$ }& \multicolumn{2}{c|}{{$B = [0.87, 0.5, 0]$, $\chi=-150\degree$}} & \multicolumn{2}{c|}{{$B = [0.71, 0.71, 0]$, $\chi=-135\degree$}} 
          \\  \hline
          
         \thead{ $M/a_0$}   &  \thead{$\left(\frac{\text{max}\,\Delta \text{I}}{\text{I}_\text{Sch}} \, 10^{-4}, \, \phi \, \right)$} & \thead{$\left(\frac{\text{max}\,\Delta \text{EVPA}}{\text{EVPA}_\text{Sch}} \, 10^{-5} , \, \phi \, \right)$} &  \thead{$\left(\frac{\text{max}\,\Delta \text{I}}{\text{I}_\text{Sch}} \, 10^{-4}, \, \phi \, \right)$} & \thead{$\left(\frac{\text{max}\,\Delta \text{EVPA}}{\text{EVPA}_\text{Sch}} \, 10^{-5} , \, \phi \, \right)$} 
          \\  \hline
          
           \thead{$\vspace{0.1mm}\text{$10^{-3}$}$\vspace{0.1mm}}  &  \thead{(75.238, 0.2$\pi$)} & \thead{(28.247, 1.12$\pi$)} &  \thead{(71.403, 0.16$\pi$)} & \thead{(40.272, 1.00$\pi$)} 
          \\  \hline

          \thead{\vspace{0.1mm}$\text{$10^{-4}$}$\vspace{0.1mm}} &  \thead{(7.543, 0.2$\pi$)} & \thead{(2.819, 1.12$\pi$)} &  \thead{(7.156, 0.16$\pi$)} & \thead{(4.019, 1.00$\pi$)} 
          \\  \hline

          \thead{\vspace{0.1mm}$\text{$10^{-5}$}$\vspace{0.1mm}}  &  \thead{(0.754, 0.2$\pi$)} & \thead{(0.282, 1.12$\pi$)} &  \thead{(0.714, 0.16$\pi$)} & \thead{(0.401, 1.00$\pi$)} 
          \\  \hline
          
        \thead{\vspace{0.1mm}$\text{$10^{-6}$}$\vspace{0.1mm}}  &  \thead{(0.075, 0.2$\pi$)} & \thead{(0.028, 1.12$\pi$)} &  \thead{(0.069, 0.16$\pi$)} & \thead{(0.039, 1.00$\pi$)} 
          \\  \hline
          
        \thead{\vspace{0.1mm}$\text{$10^{-7}$}$\vspace{0.1mm}}  &  \thead{(0.007, 0.2$\pi$)} & \thead{(0.003, 1.12$\pi$)} &  \thead{(0.006, 0.16$\pi$)} & \thead{(0.003, 1.00$\pi$)} 
          \\  \hline

       \end{tabular}
      \caption{Deviation of the linear polarization intensity and direction for  black holes with a dark matter halo from the Schwarzschild black hole. In each case we give the maximum relative deviations $\text{max}\,\Delta\text{I}/\text{I}_\text{Sch}$ and $\text{max}\,\Delta\text{EVPA}/\text{EVPA}_\text{Sch}$  with respect to the Schwarzschild solution, which are reached in the polarized images, and the corresponding azimuthal angle. The emission radius for the Schwarzschild spacetime is $r=4M$. }
    \label{table:theta_20_1}
\end{table}

\begin{table}[h!]
    \centering
      \begin{tabular}{|c|c|c|c|c|}
       \hline
       \thead{ $i = 20^{\circ}$ }& \multicolumn{2}{c|}{{$B = [0.5, 0.87, 0]$, $\chi=-120\degree$}} & \multicolumn{2}{c|}{{$B = [0, 0, 1]$, $\chi=-150\degree$}} 
          \\  \hline
          
         \thead{ $M/a_0$}   &  \thead{$\left(\frac{\text{max}\,\Delta \text{I}}{\text{I}_\text{Sch}} \, 10^{-4}, \, \phi \, \right)$} & \thead{$\left(\frac{\text{max}\,\Delta \text{EVPA}}{\text{EVPA}_\text{Sch}} \, 10^{-5} , \, \phi \, \right)$} &  \thead{$\left(\frac{\text{max}\,\Delta \text{I}}{\text{I}_\text{Sch}} \, 10^{-4}, \, \phi \, \right)$} & \thead{$\left(\frac{\text{max}\,\Delta \text{EVPA}}{\text{EVPA}_\text{Sch}} \, 10^{-5} , \, \phi \, \right)$} 
          \\  \hline
          
           \thead{$\vspace{0.1mm}\text{$10^{-3}$}$\vspace{0.1mm}}  &  \thead{(67.050, 0.08$\pi$)} & \thead{(35.207, 0.962$\pi$)} &  \thead{(54.022, 1.68$\pi$)} & \thead{(128.023, 0.68$\pi$)} 
          \\  \hline

          \thead{\vspace{0.1mm}$\text{$10^{-4}$}$\vspace{0.1mm}} &  \thead{(6.720, 0.08$\pi$)} & \thead{(3.515, 0.962$\pi$)} &  \thead{(5.410, 1.68$\pi$)} & \thead{(12.795, 0.68$\pi$)} 
          \\  \hline

          \thead{\vspace{0.1mm}$\text{$10^{-5}$}$\vspace{0.1mm}}  &  \thead{(0.672, 0.08$\pi$)} & \thead{(0.351, 0.962$\pi$)} &  \thead{(0.539, 1.68$\pi$)} & \thead{(1.276, 0.68$\pi$)} 
          \\  \hline
          
        \thead{\vspace{0.1mm}$\text{$10^{-6}$}$\vspace{0.1mm}}  & \thead{(0.068, 0.08$\pi$)} & \thead{(0.035, 0.962$\pi$)} &  \thead{(0.052, 1.68$\pi$)} & \thead{(0.124, 0.68$\pi$)} 
          \\  \hline
          
        \thead{\vspace{0.1mm}$\text{$10^{-7}$}$\vspace{0.1mm}}  & \thead{(0.006, 0.08$\pi$)} & \thead{(0.003, 0.962$\pi$)}  &  \thead{(0.005, 1.68$\pi$)} & \thead{(0.011, 0.68$\pi$)} 
          \\  \hline

       \end{tabular}
      \caption{Deviation of the linear polarization intensity and direction for black holes with a dark matter halo from the Schwarzschild black hole. In each case we give the maximum relative deviations $\text{max}\,\Delta\text{I}/\text{I}_\text{Sch}$ and $\text{max}\,\Delta\text{EVPA}/\text{EVPA}_\text{Sch}$  with respect to the Schwarzschild solution, which are reached in the polarized images, and the corresponding azimuthal angle. The emission radius for the Schwarzschild spacetime is $r=4M$. }
    \label{table:theta_20_2}
\end{table}

Based on these results, we conclude that at small inclination angles, the solution describing a black hole with a dark matter halo can reproduce the overall morphology of the polarized images characteristic for the Schwarzschild black hole. Given an appropriate magnetic field configuration, the polarization pattern remains consistent with the observational data from M87*. To determine whether it is possible to distinguish between classical black holes and those surrounded by dark matter using polarimetric observations, we proceed with a quantitative comparison of the differences in intensity and the rotation angle of the polarization vector. These comparisons should be made at the same apparent image size and with morphological features consistent with those of an orbit around the Schwarzschild black hole, rather than solely at equal emission radii. Although the emission sources are located at different radial positions, the resulting image differences remain negligible for astrophysically realistic values of halo compactness $M/a_{0}$, since both types of spacetimes exhibit similar light-bending and focusing effects in direct and indirect images. An exception arises at high compactness, where the increased presence of dark matter in the halo leads to significantly larger image sizes compared to those typical of the Schwarzschild black hole. 

\begin{figure}[t!]
    \centering
    \includegraphics[width=14.2cm]{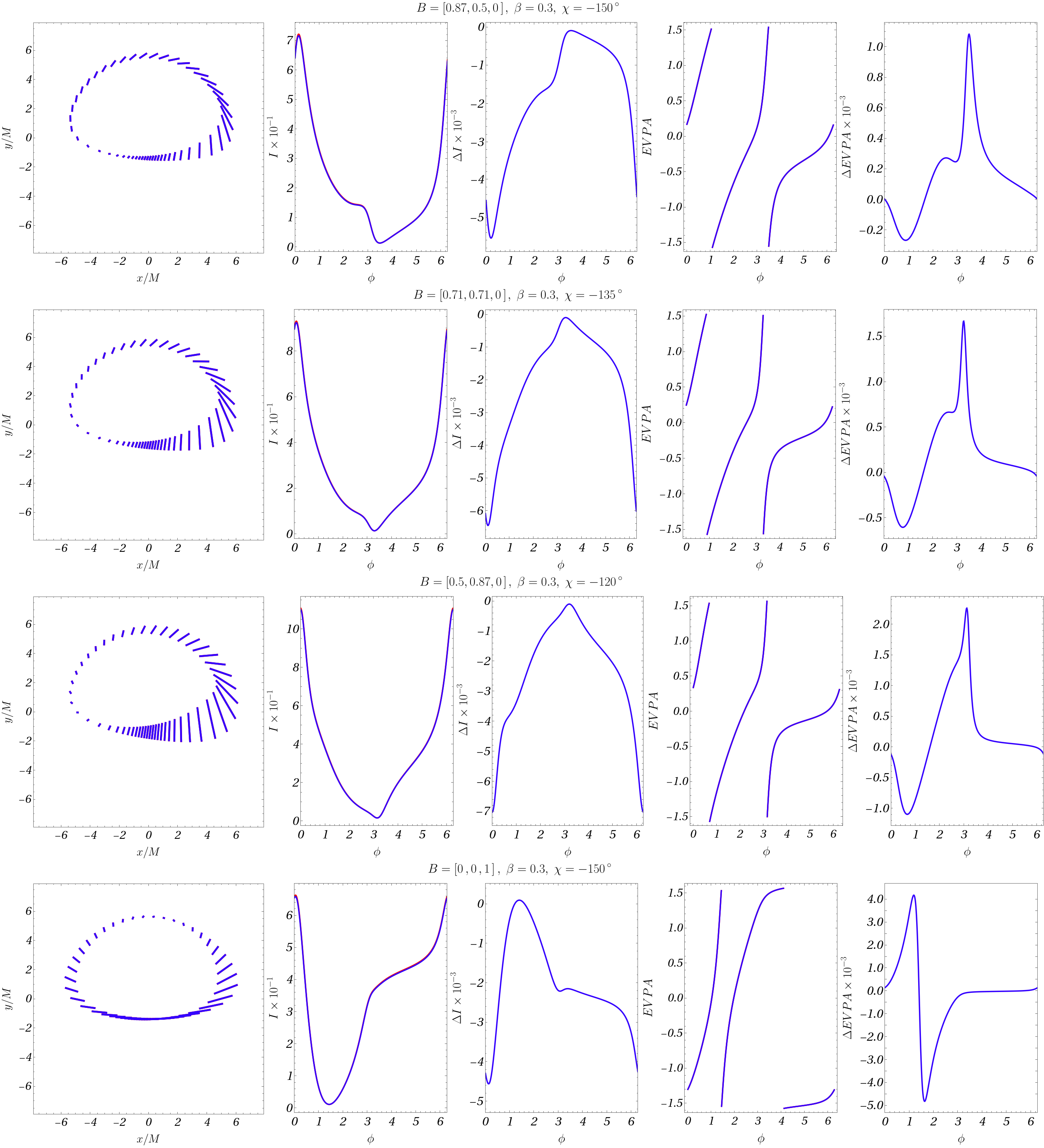}
    \caption{Linear polarization for a black hole with a dark matter halo with compactness parameter $C=10^{-3}$ (blue line) compared to the Schwarzschild black hole (red line). The inclination angle is $i = 70^\circ$ and the emission radius for the Schwarzschild spacetime is $r=4M$. We analyze the polarization intensity I and direction EVPA, and their deviation from the Schwarzschild black hole $\Delta \text{I}$ and $\Delta \text{EVPA}$ (see main text).}
    \label{fig:direct-inclination70}
\end{figure}

\begin{table}[h!]
    \centering
      \begin{tabular}{|c|c|c|c|c|}
       \hline
       \thead{ $i = 70^{\circ}$ }& \multicolumn{2}{c|}{{$B = [0.87, 0.5, 0]$, $\chi=-150\degree$}} & \multicolumn{2}{c|}{{$B = [0.71, 0.71, 0]$, $\chi=-135\degree$}} 
          \\  \hline
          
        \thead{ $M/a{_o}$ }  & \thead{$\left(\frac{\text{max}\,\Delta \text{I}}{\text{I}_\text{Sch}} \, 10^{-4}, \, \phi \, \right)$} & \thead{$\left(\frac{\text{max}\,\Delta \text{EVPA}}{\text{EVPA}_\text{Sch}} \, 10^{-5} , \, \phi \, \right)$} & \thead{$\left(\frac{\text{max}\,\Delta \text{I}}{\text{I}_\text{Sch}} \, 10^{-4}, \, \phi \, \right)$} & \thead{$\left(\frac{\text{max}\,\Delta \text{EVPA}}{\text{EVPA}_\text{Sch}} \, 10^{-5} , \, \phi \, \right)$} 
          \\  \hline

           \thead{$\vspace{0.1mm}\text{$10^{-3}$}$\vspace{0.1mm}}  &  \thead{(79.314, 0.08$\pi$)} & \thead{(70,850, 1.12$\pi$)} &  \thead{(69.922, 0.04$\pi$)} & \thead{(120.333, 1.04$\pi$)} 
          \\  \hline

          \thead{\vspace{0.1mm}$\text{$10^{-4}$}$\vspace{0.1mm}} &  \thead{(7.951, 0.08$\pi$)} & \thead{(7.075, 1.12$\pi$)} &  \thead{(7.006, 0.04$\pi$)} & \thead{(12.010, 1.04$\pi$)} 
          \\  \hline

          \thead{\vspace{0.1mm}$\text{$10^{-5}$}$\vspace{0.1mm}}  &  \thead{(0.795, 0.08$\pi$)} & \thead{(0.707, 1.12$\pi$)} &  \thead{(0.699, 0.04$\pi$)} & \thead{(1.197, 1.04$\pi$)} 
          \\  \hline
          
        \thead{\vspace{0.1mm}$\text{$10^{-6}$}$\vspace{0.1mm}}  &  \thead{(0.079, 0.08$\pi$)} & \thead{(0,070, 1.12$\pi$)} &  \thead{(0.068, 0.04$\pi$)} & \thead{(0.116, 1.04$\pi$)} 
          \\  \hline
          
        \thead{\vspace{0.1mm}$\text{$10^{-7}$}$\vspace{0.1mm}}  &  \thead{(0.008, 0.08$\pi$)} & \thead{(0,007, 1.12$\pi$)} &  \thead{(0.006, 0.04$\pi$)} & \thead{(0.010, 1.04$\pi$)} 
          \\  \hline

       \end{tabular}
      \caption{Deviation of the linear polarization intensity and direction for for black holes with a dark matter halo from the Schwarzschild black hole. In each case we give the maximum relative deviations $\text{max}\,\Delta\text{I}/\text{I}_\text{Sch}$ and $\text{max}\,\Delta\text{EVPA}/\text{EVPA}_\text{Sch}$  with respect to the Schwarzschild solution, which are reached in the polarized images, and the corresponding azimuthal angle. The emission radius for the Schwarzschild spacetime is $r=4M$. }
    \label{table:theta_70_1}
\end{table}

\begin{table}[h!]
    \centering
      \begin{tabular}{|c|c|c|c|c|}
       \hline
       \thead{ $i = 70^{\circ}$ }& \multicolumn{2}{c|}{{$B = [0.5, 0.87, 0]$, $\chi=-120\degree$}} & \multicolumn{2}{c|}{{$B = [0, 0, 1]$, $\chi=-150\degree$}} 
          \\  \hline
          
        \thead{ $M/a{_o}$ }  & \thead{$\left(\frac{\text{max}\,\Delta \text{I}}{\text{I}_\text{Sch}} \, 10^{-4}, \, \phi \, \right)$} & \thead{$\left(\frac{\text{max}\,\Delta \text{EVPA}}{\text{EVPA}_\text{Sch}} \, 10^{-5} , \, \phi \, \right)$} & \thead{$\left(\frac{\text{max}\,\Delta \text{I}}{\text{I}_\text{Sch}} \, 10^{-4}, \, \phi \, \right)$} & \thead{$\left(\frac{\text{max}\,\Delta \text{EVPA}}{\text{EVPA}_\text{Sch}} \, 10^{-4} , \, \phi \, \right)$} 
          \\  \hline

           \thead{$\vspace{0.1mm}\text{$10^{-3}$}$\vspace{0.1mm}}  &  \thead{(63,339, 0)} & \thead{(143,845, 1.002$\pi$)} &  \thead{(70.406,  0.04$\pi$)} & \thead{(66.366, 0.52$\pi$)} 
          \\  \hline

          \thead{\vspace{0.1mm}$\text{$10^{-4}$}$\vspace{0.1mm}} &  \thead{(6.347, 0)} & \thead{(14.365, 1.002$\pi$)} &  \thead{(7.055,  0.04$\pi$)} & \thead{(6.653, 0.52$\pi$)} 
          \\  \hline

          \thead{\vspace{0.1mm}$\text{$10^{-5}$}$\vspace{0.1mm}}  &  \thead{(0.634, 0)} & \thead{(1.435, 1.002$\pi$)} &  \thead{(0.704,  0.04$\pi$)} & \thead{(0.663, 0.52$\pi$)} 
          \\  \hline
          
        \thead{\vspace{0.1mm}$\text{$10^{-6}$}$\vspace{0.1mm}}  &  \thead{(0.063, 0)} & \thead{(0.143, 1.002$\pi$)} &  \thead{(0.068,  0.04$\pi$)} & \thead{(0.065, 0.52$\pi$)} 
          \\  \hline
          
        \thead{\vspace{0.1mm}$\text{$10^{-7}$}$\vspace{0.1mm}}  &  \thead{(0.006, 0)} & \thead{(0.014, 1.002$\pi$)} &  \thead{(0.006,  0.04$\pi$)} & \thead{(0.006, 0.52$\pi$)} 
          \\  \hline

       \end{tabular}
      \caption{Deviation of the linear polarization intensity and direction for for black holes with a dark matter halo from the Schwarzschild black hole. In each case we give the maximum relative deviations $\text{max}\,\Delta\text{I}/\text{I}_\text{Sch}$ and $\text{max}\,\Delta\text{EVPA}/\text{EVPA}_\text{Sch}$  with respect to the Schwarzschild solution, which are reached in the polarized images, and the corresponding azimuthal angle. The emission radius for the Schwarzschild spacetime is $r=4M$. }
    \label{table:theta_70_2}
\end{table}

The results of our analysis are presented in Figure \ref{fig:direct-inclination20}, where we examine the polarization characteristics of direct images formed at the observable location of an orbit with radius $r_{s}=4M$ for a Schwarzschild black hole, in comparison with the solution describing a black hole surrounded by a dark matter halo. For a specific value of the compactness parameter, $M/a_{0}=10^{-3}$, which characterizes the dark matter content in the galactic halo, we analyze the variation of the polarization intensity and direction as a function of the azimuthal coordinate in the resulting images. The deviations from the Schwarzschild solution are evaluated through the relative intensity difference $\Delta \text{I} = \text{I}_\text{DM} - \text{I}_\text{Sch}$ and the difference in the orientation angle of the polarization vector $\Delta \text{EVPA} =\text{ EVPA}_\text{DM} - \text{EVPA}_\text{Sch}$ at each point of the image. The results show that regardless of the value of the compactness parameter, the profiles of the intensity and its orientation follow the same dependence on the azimuthal angle as observed for the Schwarzschild black hole. The positions of the peaks and minima in the distribution also coincide. Since the Schwarzschild geometry corresponds to the limiting case of the solution with a dark matter halo when the compactness tends to zero, the polarization properties of the black hole surrounded by dark matter can become indistinguishably close to those of a pure Schwarzschild black hole. The deviations increase significantly with higher values of $M/a_{0}$, reaching their maximum for increasingly compact halos. In Table \ref{table:theta_20_1}, we present the maximum deviations in polarization intensity and direction for various values of the compactness parameter under different equatorial magnetic field configurations. These results provide an estimate for the largest possible deviations in the polarization properties of a black hole surrounded by a dark matter halo.

\begin{figure}[t!]
    \centering
    \includegraphics[width=14.2cm]{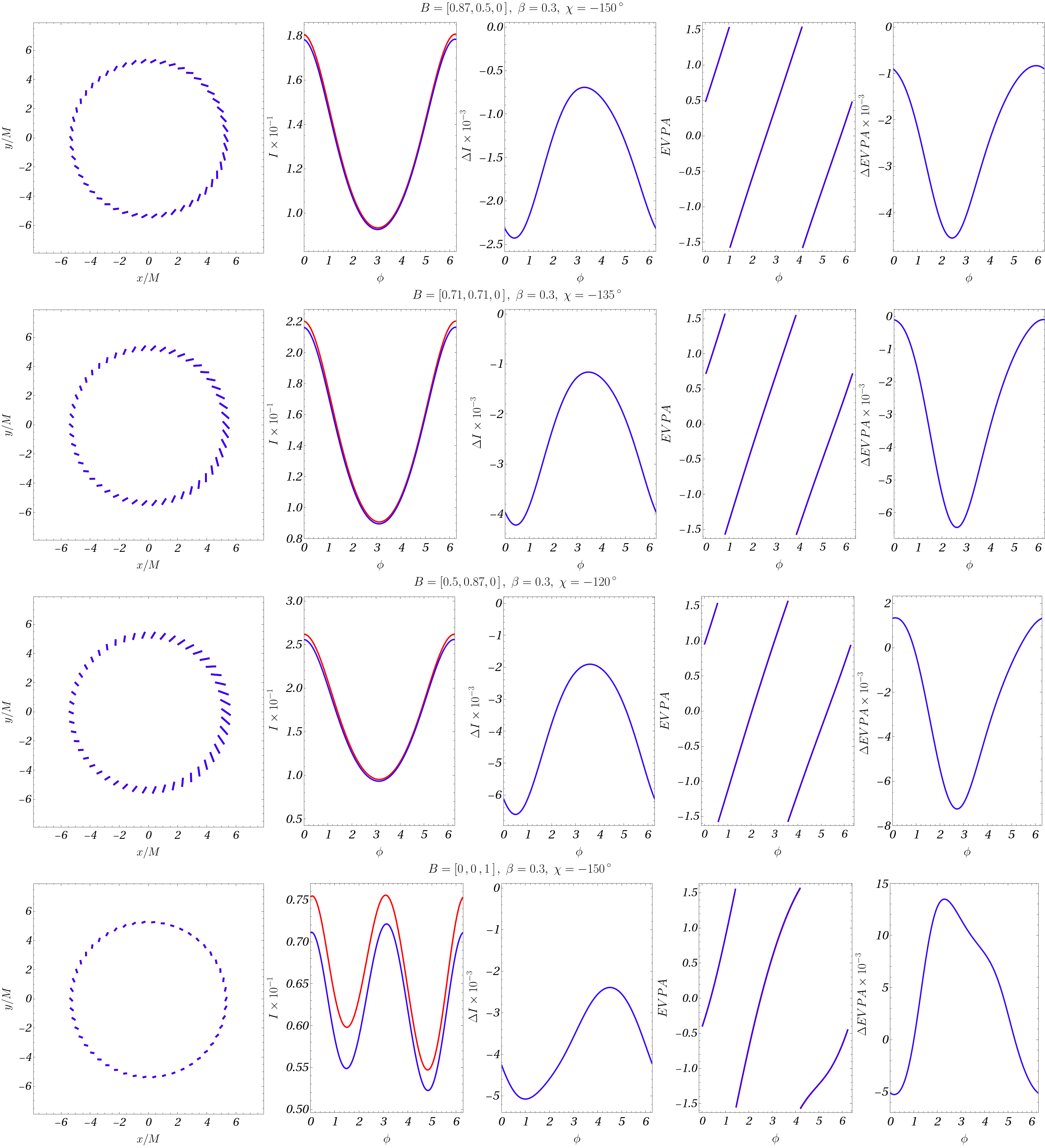}
    \caption{Polarization of the indirect images of order $k = 1$ for a black hole with a dark matter halo with compactness parameter $C=10^{-3}$ (blue line) compared to the Schwarzschild black hole (red line). The inclination angle is $i = 20^\circ$ and the emission radius for the Schwarzschild spacetime is $r=4M$. We analyze the polarization intensity I and direction EVPA, and their deviation from the Schwarzschild black hole $\Delta \text{I}$ and $\Delta \text{EVPA}$ (see main text).}
    \label{fig:indirect-inclination20}
\end{figure}

\begin{table}[h!]
    \centering
      \begin{tabular}{|c|c|c|c|c|}
       \hline
       \thead{ $i = 20^{\circ}$ }& \multicolumn{2}{c |}{{$B = [0.87, 0.5, 0]$, $\chi=-150\degree$}} & \multicolumn{2}{c|}{{$B = [0.71, 0.71, 0]$, $\chi=-135\degree$}} 
          \\  \hline
          
         \thead{ $M/a_0$}   &  \thead{$\left(\frac{\text{max}\,\Delta \text{I}}{\text{I}_\text{Sch}} \, 10^{-4}, \, \phi \,\right)$} & \thead{$\left(\frac{\text{max}\,\Delta \text{EVPA}}{\text{EVPA}_\text{Sch}} \, 10^{-4} , \, \phi \, \right)$} &  \thead{$\left(\frac{\text{max}\,\Delta \text{I}}{\text{I}_\text{Sch}} \, 10^{-4}, \, \phi \, \right)$} & \thead{$\left(\frac{\text{max}\,\Delta \text{EVPA}}{\text{EVPA}_\text{Sch}} \, 10^{-4} , \, \phi \, \right)$} 
          \\  \hline
          
           \thead{$\vspace{0.1mm}\text{$10^{-3}$}$\vspace{0.1mm}}  &  \thead{(139.208, 0.12$\pi$)} & \thead{(241.560, 0.76$\pi$)} &  \thead{(204.246, $0.16\pi$)} & \thead{(180.512, 0.84$\pi$)} 
          \\  \hline

          \thead{\vspace{0.1mm}$\text{$10^{-4}$}$\vspace{0.1mm}} &  \thead{(13.795, 0.12$\pi$)} & \thead{(24.517, 0.76$\pi$)} &  \thead{(20.372, 0.16$\pi$)} & \thead{(18.205, 0.84$\pi$)} 
          \\  \hline

          \thead{\vspace{0.1mm}$\text{$10^{-5}$}$\vspace{0.1mm}}  &  \thead{(1.377
, 0.12$\pi$)} & \thead{(2.454, 0.76$\pi$)} &  \thead{(2.031, 0.16$\pi$)} & \thead{(1.816, 0.84$\pi$)} 
          \\  \hline
          
        \thead{\vspace{0.1mm}$\text{$10^{-6}$}$\vspace{0.1mm}}  &  \thead{(0.137, 0.12$\pi$)} & \thead{(0.244, 0.76$\pi$)} &  \thead{(0.198, 0.16$\pi$)} & \thead{(0.177, 0.84$\pi$)} 
          \\  \hline
          
        \thead{\vspace{0.1mm}$\text{$10^{-7}$}$\vspace{0.1mm}}  &  \thead{(0.013, 0.12$\pi$)} & \thead{(0.023, 0.76$\pi$)} &  \thead{(0.017, 0.16$\pi$)} & \thead{(0.015, 0.84$\pi$)} 
          \\  \hline

       \end{tabular}
      \caption{Deviation of the polarization intensity and EVPA for the indirect images of order $k = 1$ for the black holes with a dark matter halo from the Schwarzschild black hole. In each case we give the maximum relative deviations $\text{max}\,\Delta\text{I}/\text{I}_\text{Sch}$ and $\text{max}\,\Delta\text{EVPA}/\text{EVPA}_\text{Sch}$  with respect to the Schwarzschild solution, which are reached in the polarized images, and the corresponding azimuthal angle.  The velocity of the fluid is $\beta = 0.3$.  The emission radius for the Schwarzschild spacetime is $r=4M{_{\rm BH}}$. }
    \label{table:theta_20_1_indirect}
\end{table}

\begin{table}[h!]
    \centering
      \begin{tabular}{|c|c|c|c|c|}
       \hline
       \thead{ $i = 20^{\circ}$ }& \multicolumn{2}{c |}{{$B = [0.5, 0.87, 0]$, $\chi=-120\degree$}} & \multicolumn{2}{c|}{{$B = [0, 0, 1]$, $\chi=-150\degree$}} 
          \\  \hline
          
         \thead{ $M/a_0$}   &  \thead{$\left(\frac{\text{max}\,\Delta \text{I}}{\text{I}_\text{Sch}} \, 10^{-4}, \, \phi \,\right)$} & \thead{$\left(\frac{\text{max}\,\Delta \text{EVPA}}{\text{EVPA}_\text{Sch}} \, 10^{-4} , \, \phi \, \right)$} &  \thead{$\left(\frac{\text{max}\,\Delta \text{I}}{\text{I}_\text{Sch}} \, 10^{-4}, \, \phi \, \right)$} & \thead{$\left(\frac{\text{max}\,\Delta \text{EVPA}}{\text{EVPA}_\text{Sch}} \, 10^{-4} , \, \phi \, \right)$} 
          \\  \hline
          
           \thead{$\vspace{0.1mm}\text{$10^{-3}$}$\vspace{0.1mm}}  &  \thead{(269.111, 0.16$\pi$)} & \thead{(93.094, 0.88$\pi$)} &  \thead{(800.029, 0.32$\pi$)} & \thead{(432.976, 0.72$\pi$)}
          \\  \hline

          \thead{\vspace{0.1mm}$\text{$10^{-4}$}$\vspace{0.1mm}} &  \thead{(26.988, 0.16$\pi$)} & \thead{(11.185, 0.84$\pi$)} &  \thead{(81.501, 0.32$\pi$)} & \thead{(42.125, 0.72$\pi$)} 
          \\  \hline

          \thead{\vspace{0.1mm}$\text{$10^{-5}$}$\vspace{0.1mm}}  &  \thead{(2.698, 0.16$\pi$)} & \thead{(1.118, 0.84$\pi$)} &  \thead{(8.140, 0.32$\pi$)} & \thead{(4.188, 0.72$\pi$)}
          \\  \hline
          
        \thead{\vspace{0.1mm}$\text{$10^{-6}$}$\vspace{0.1mm}}  & \thead{(0.268, 0.16$\pi$)} & \thead{(0.111, 0.84$\pi$)} &  \thead{(0.792, 0.32$\pi$)} & \thead{(0.407, 0.72$\pi$)}
          \\  \hline
          
        \thead{\vspace{0.1mm}$\text{$10^{-7}$}$\vspace{0.1mm}}  & \thead{(0.025, 0.16$\pi$)} & \thead{(0.011, 0.84$\pi$)}  &  \thead{(0.069, 0.32$\pi$)} & \thead{(0.035, 0.72$\pi$)}
          \\  \hline

       \end{tabular}
      \caption{Deviation of the polarization intensity and direction for the indirect images of order $k = 1$ for the black holes with a dark matter halo from the Schwarzschild black hole. In each case we give the maximum relative deviation $\text{max}\,\Delta\text{I}/\text{I}_\text{Sch}$ and $\text{max}\,\Delta\text{EVPA}/\text{EVPA}_\text{Sch}$  with respect to the Schwarzschild solution, which are reached in the polarized images, and the corresponding azimuthal angle.  The velocity of the fluid is $\beta = 0.3$.  The emission radius for the Schwarzschild spacetime is $r=4M{_{\rm BH}}$. }
    \label{table:theta_20_2_indirect}
\end{table}

In configurations with an equatorially aligned magnetic field, we additionally observe that the extent of the deviation in the polarization profiles between a black hole surrounded by a dark matter halo and the Schwarzschild black hole is influenced by the specific orientation of the magnetic field. As the radial component of the magnetic field increases, the difference in polarization intensity becomes more pronounced, whereas the twist of the polarization vector increasingly resembles that in the Schwarzschild geometry. Across all considered values of the compactness parameter, the deviations remain well below the threshold of detectability at the current angular resolution of the EHT. Specifically, the deviation in polarization intensity does not exceed $0.75 \%$ for any of the magnetic field configurations studied, and the change in the polarization vector orientation remains under $0.13\%$.

The solutions describing a black hole surrounded by a dark matter halo reproduce an almost identical twist of the polarization vector across the image (max $\Delta \text{EVPA} / \text{EVPA}_{\text{Sch}}$ $ < 0.03\%$) for all considered values of the compactness parameter, with the deviation further decreasing for smaller halo compactness. This behavior is most pronounced for a magnetic field direction $ B = [0.87, 0.5, 0] $, which provides the best agreement with the observed polarized images of M87* when the spacetime is modeled by the Schwarzschild geometry.

\begin{figure}[t!]
    \centering
    \includegraphics[width=14.2cm]{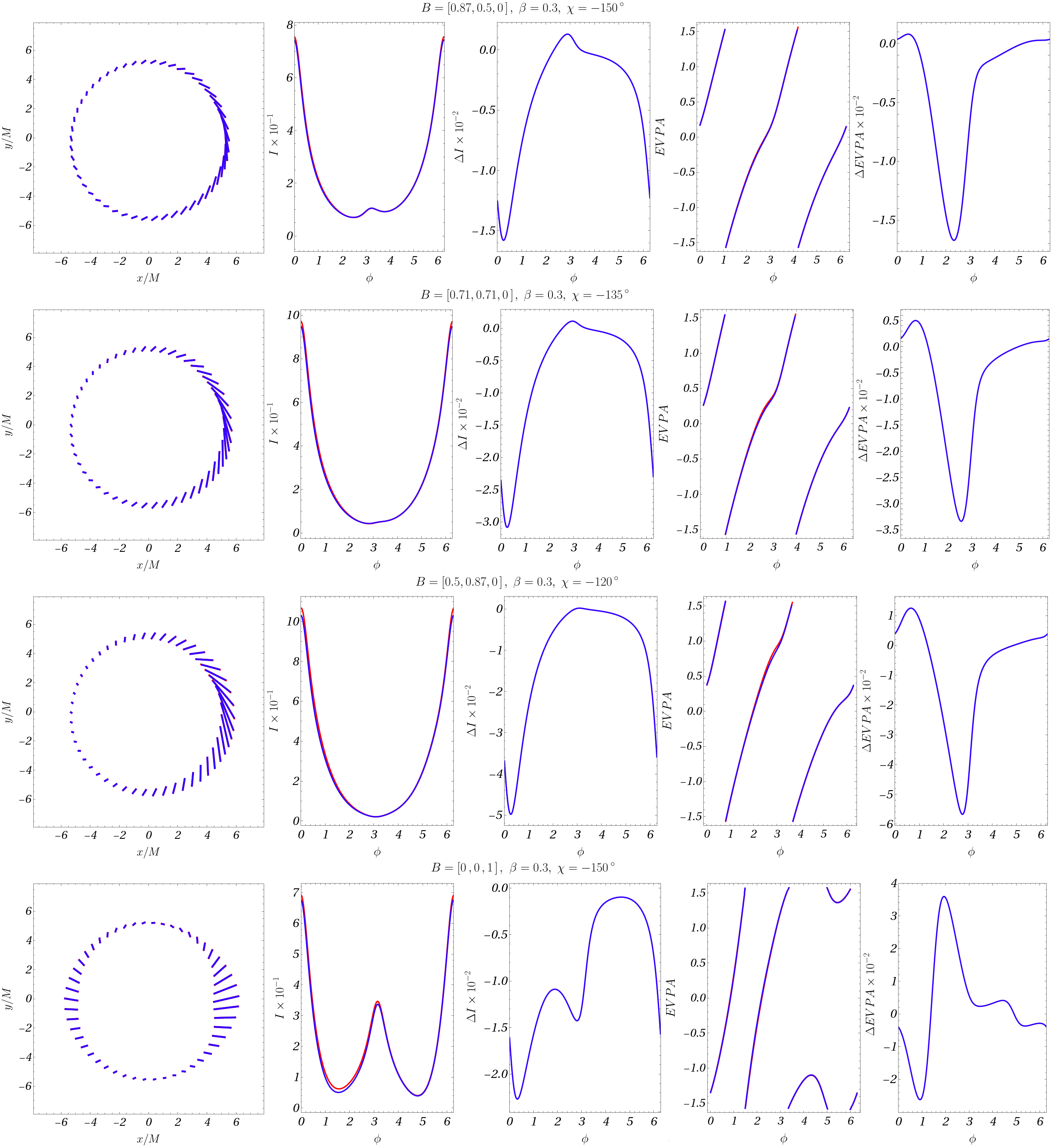}
    \caption{Polarization of the indirect images of order $k = 1$ for a black hole with a dark matter halo with compactness parameter $C=10^{-3}$ (blue line) compared to the Schwarzschild black hole (red line). The inclination angle is $i = 70^\circ$ and the emission radius for the Schwarzschild spacetime is $r=4M$. We analyze the polarization intensity I and direction EVPA, and their deviation from the Schwarzschild black hole $\Delta \text{I}$ and $\Delta \text{EVPA}$ (see main text).}
    \label{fig:indirect-inclination70}
\end{figure}

To extend the scope of our analysis, we explore how the large angle of light deviations impacts the polarization characteristics of direct emission, which is realized particularly at high viewing angles. Although these high-inclination configurations are not intended to model M87*, they offer valuable insights from a theoretical standpoint and may prove relevant for future EHT observations targeting other galactic sources. For consistency, the simulations follow the same framework as those in Figure \ref{fig:direct-inclination20}, but are adapted to large inclination angles.

The results of this study are presented in Figure \ref{fig:direct-inclination70}, which illustrates the polarization at the apparent location of the orbit with radius $ r_s = 4M $ in the spacetime of a black hole surrounded by a dark matter halo, viewed at an inclination angle of $ i = 70^\circ $. The overall structure of the polarization pattern remains qualitatively similar to the Schwarzschild case; however, the magnitude of the observed deviations increases with higher inclination. Table \ref{table:theta_70_1} provides a quantitative summary of these differences, showing the maximum deviations for various values of the compactness parameter and magnetic field orientations. The largest discrepancy in polarization intensity, reaching up to $ 0.79\% $, occurs for the magnetic field vector $ B = [0.87, 0.5, 0] $, where the radial component is dominant. In contrast, the deviation in the orientation of the polarization vector (EVPA) follows the opposite trend. The most significant difference, measured at $ 0.14\% $, is found when the radial component is reduced, specifically for $ B = [0.5, 0.87, 0] $.

Finally, the configuration with a vertically oriented magnetic field exhibits a symmetric polarization pattern, where the left and right regions of the image display the strongest polarization, while the upper and lower parts are significantly weaker. The relative deviations in both polarization intensity and the rotation angle of the polarization vector fall within intermediate ranges compared to the maximum deviations discussed above, as highlighted in Tables \ref{table:theta_20_2} and \ref{table:theta_70_2}.

\subsection{Indirect images}

In this section, we examine the polarization properties of indirect images formed by null geodesics that complete half an orbit around a black hole surrounded by a dark matter halo before reaching the observer. In this case, the azimuthal coordinate of the photons varies within the range $\phi \in [0, (k+1)\pi]$, where for the indirect images considered here $k = 1$, denoting the image order. Such images provide additional information about the gravitational field of the central object, as the photons experience a greater time delay while propagating near the event horizon, unlike those forming direct images. 

To investigate this effect, we compute the observable polarization emitted from orbits with a fixed radius, assuming a compactness parameter of the halo $M/a_{0} = 10^{-3}$, both at low and high observer inclination angles. A comparison with the Schwarzschild black hole is made by simulating the polarization at the apparent location of the orbit at $r_{s} = 4M$, generated by geodesic trajectories of order $k = 1$ in both spacetime geometries (see Figures \ref{fig:indirect-inclination20} and \ref{fig:indirect-inclination70}).

The results for an observer at inclination angle $i = 20^\circ$ are shown in Figure \ref{fig:indirect-inclination20}, where we compare the polarization profiles of a black hole with a dark matter halo and the Schwarzschild black hole. We find that the deviations in polarization intensity are more pronounced than in the case of direct images. For instance, with a magnetic field configuration $B = [0.5, 0.87, 0]$, the polarized flux for the black hole with a dark matter halo is $2.7\%$ lower than for the Schwarzschild case. Even in the case of a purely vertical magnetic field, $B = [0, 0, 1]$, the intensity is reduced by a factor of 14.8 compared to the direct image, with deviations due to the presence of dark matter reaching about $8\%$, as summarized in Table \ref{table:theta_20_2_indirect}.

\begin{table}[t!]
    \centering
      \begin{tabular}{|c|c|c|c|c|}
       \hline
       \thead{ $i = 70^{\circ}$ }& \multicolumn{2}{c |}{{$\text{B = [0.87, 0.5, 0]}$, $\chi=-150\degree$}} & \multicolumn{2}{c|}{{$\text{B = [0.71, 0.71, 0]}$, $\chi=-135\degree$}} 
          \\  \hline
          
        \thead{ $M/a{_o}$ }  & \thead{$\left(\frac{\text{max}\,\Delta \text{I}}{\text{I}_\text{Sch}} \, 10^{-4}, \, \phi \, \right)$} & \thead{$\left(\frac{\text{max}\,\Delta \text{EVPA}}{\text{EVPA}_\text{Sch}} \, 10^{-4} , \, \phi \, \right)$} & \thead{$\left(\frac{\text{max}\,\Delta \text{I}}{\text{I}_\text{Sch}} \, 10^{-4}, \, \phi \, \right)$} & \thead{$\left(\frac{\text{max}\,\Delta \text{EVPA}}{\text{EVPA}_\text{Sch}} \, 10^{-3} , \, \phi \, \right)$} 
          \\  \hline

           \thead{$\vspace{0.1mm}\text{$10^{-3}$}$\vspace{0.1mm}}  &  \thead{(263.659, 0.08$\pi$)} & \thead{(471.957, 0.72$\pi$)} &  \thead{(394.290, 0.08$\pi$)} & \thead{(198.083, 0.80$\pi$)} 
          \\  \hline

          \thead{\vspace{0.1mm}$\text{$10^{-4}$}$\vspace{0.1mm}} &  \thead{(25.895, 0.08$\pi$)} & \thead{(48.878, 0.72$\pi$)} &  \thead{(39.056, 0.08$\pi$)} & \thead{(20.356, 0.80$\pi$)} 
          \\  \hline

          \thead{\vspace{0.1mm}$\text{$10^{-5}$}$\vspace{0.1mm}}  &  \thead{(2.583, 0.08$\pi$)} & \thead{(4.902, 0.72$\pi$)} &  \thead{(3.890, 0.08$\pi$)} & \thead{(2.035, 0.80$\pi$)} 
          \\  \hline
          
        \thead{\vspace{0.1mm}$\text{$10^{-6}$}$\vspace{0.1mm}}  &  \thead{(0.256, 0.08$\pi$)} & \thead{(0.487, 0.72$\pi$)} &  \thead{(0.378, 0.08$\pi$)} & \thead{(0.198, 0.80$\pi$)} 
          \\  \hline
          
        \thead{\vspace{0.1mm}$\text{$10^{-7}$}$\vspace{0.1mm}}  &  \thead{(0.024, 0.08$\pi$)} & \thead{(0.046, 0.72$\pi$)} &  \thead{(0.033, 0.08$\pi$)} & \thead{(0.017, 0.80$\pi$)} 
          \\  \hline

       \end{tabular}
      \caption{Deviation of the polarization intensity and direction for the indirect images of order $k = 1$ for the black holes with a dark matter halo from the Schwarzschild black hole. In each case we give the maximum relative deviations $\text{max}\,\Delta\text{I}/\text{I}_\text{Sch}$ and $\text{max}\,\Delta\text{EVPA}/\text{EVPA}_\text{Sch}$  with respect to the Schwarzschild solution, which are reached in the polarized images, and the corresponding azimuthal angle.  The velocity of the fluid is $\beta = 0.3$.  The emission radius for the Schwarzschild spacetime is $r=4M{_{\rm BH}}$. }
    \label{table:theta_70_1_indirect}
\end{table}

\begin{table}[t!]
    \centering
      \begin{tabular}{|c|c|c|c|c|}
       \hline
       \thead{ $i = 70^{\circ}$ }& \multicolumn{2}{c |}{{$\text{B = [0.5, 0.87, 0]}$, $\chi=-120\degree$}} & \multicolumn{2}{c|}{{$\text{B = [0, 0, 1]}$, $\chi=-150\degree$}} 
          \\  \hline
          
        \thead{ $M/a{_o}$ }  & \thead{$\left(\frac{\text{max}\,\Delta \text{I}}{\text{I}_\text{Sch}} \, 10^{-4}, \, \phi \, \right)$} & \thead{$\left(\frac{\text{max}\,\Delta \text{EVPA}}{\text{EVPA}_\text{Sch}} \, 10^{-4} , \, \phi \, \right)$} & \thead{$\left(\frac{\text{max}\,\Delta \text{I}}{\text{I}_\text{Sch}} \, 10^{-4}, \, \phi \, \right)$} & \thead{$\left(\frac{\text{max}\,\Delta \text{EVPA}}{\text{EVPA}_\text{Sch}} \, 10^{-4} , \, \phi \, \right)$} 
          \\  \hline

           \thead{$\vspace{0.1mm}\text{$10^{-3}$}$\vspace{0.1mm}}  &  \thead{(561.812, 0.08$\pi$)} & \thead{(713.086, 0.88$\pi$)} &  \thead{(540.558,  0.12$\pi$)} & \thead{(418.689, 0.60$\pi$)} 
          \\  \hline

          \thead{\vspace{0.1mm}$\text{$10^{-4}$}$\vspace{0.1mm}} &  \thead{(56.192, 0.08$\pi$)} & \thead{(72.374, 0.88$\pi$)} &  \thead{(54.835,  0.12$\pi$)} & \thead{(39.270, 0.60$\pi$)} 
          \\  \hline

          \thead{\vspace{0.1mm}$\text{$10^{-5}$}$\vspace{0.1mm}}  &  \thead{(5.616, 0.08$\pi$)} & \thead{(7.243 0.88$\pi$)} &  \thead{(5.475,  0.12$\pi$)} & \thead{(3.892, 0.60$\pi$)} 
          \\  \hline
          
        \thead{\vspace{0.1mm}$\text{$10^{-6}$}$\vspace{0.1mm}}  &  \thead{(0.558, 0.08$\pi$)} & \thead{(0.720, 0.88$\pi$)} &  \thead{(0.533,  0.12$\pi$)} & \thead{(0.378, 0.60$\pi$)} 
          \\  \hline
          
        \thead{\vspace{0.1mm}$\text{$10^{-7}$}$\vspace{0.1mm}}  &  \thead{(0.053, 0.08$\pi$)} & \thead{(0.068, 0.88$\pi$)} &  \thead{(0.046,  0.12$\pi$)} & \thead{(0.033, 0.60$\pi$)} 
          \\  \hline

       \end{tabular}
      \caption{Deviation of the polarization intensity and direction for the indirect images of order $k = 1$ for the black holes with a dark matter halo from the Schwarzschild black hole. In each case we give the maximum relative deviations $\text{max}\,\Delta\text{I}/\text{I}_\text{Sch}$ and $\text{max}\,\Delta\text{EVPA}/\text{EVPA}_\text{Sch}$  with respect to the Schwarzschild solution, which are reached in the polarized images, and the corresponding azimuthal angle.  The velocity of the fluid is $\beta = 0.3$.  The emission radius for the Schwarzschild spacetime is $r=4M{_{\rm BH}}$. }
    \label{table:theta_70_2_indirect}
\end{table}

Furthermore, the structure of the field around the ring is altered. Although the overall polarization intensity decreases, the frequency of variation of the peak values along a full rotation of the emitting orbit doubles. Two characteristic maxima are observed on the left and right sides of the image, while the corresponding two minima are located at the top and bottom parts, respectively. These features suggest that higher-order images formed in vertically aligned magnetic fields may be used to distinguish black holes from naked singularities in future high-resolution observations with the EHT.

For an observer at inclination angle $i = 70^\circ$, the results are shown in Figure \ref{fig:indirect-inclination70}. In this case, the maximum deviation in polarization intensity due to the presence of dark matter occurs for the magnetic field configuration $B = [0.5, 0.87, 0]$, where the polarization level is $8.9\%$ higher than in direct images and $5.6\%$ lower than the corresponding level for a Schwarzschild black hole, as shown in Table \ref{table:theta_70_2_indirect}. In the case of a purely vertical field, $B = [0, 0, 1]$, two intensity peaks are again observed, but now the first peak on the right side of the image is more pronounced than the second one on the left. The accompanying two minima, as in the case of $i = 20^\circ$, are again located at the top and bottom parts of the image.

The deviation in the orientation of the polarization vector also increases. At small inclination angles, for $B = [0, 0, 1]$, the maximum EVPA for the black hole with a dark matter halo is approximately $4.3\%$ higher than that for the Schwarzschild case, as shown in Table \ref{table:theta_20_2_indirect}. For $B = [0.87, 0.5, 0]$, this increase is reduced by about $50\%$, as indicated in Table \ref{table:theta_20_1_indirect}. Despite these quantitative differences, the overall shape and symmetry of the polarization direction distribution remain similar to that of the Schwarzschild geometry. At higher inclination angles, the differences become even more significant. The largest deviations in the observed polarized flux due to the presence of dark matter occur for the magnetic field configuration $B = [0.5, 0.87, 0]$, where the intensity increases by about $7.1\%$ compared to the Schwarzschild case, as presented in Table \ref{table:theta_70_2_indirect}. For the configuration with the strongest radial component, $B = [0.87, 0.5, 0]$, the increase is reduced by approximately $34\%$, as summarized in Table \ref{table:theta_70_1_indirect}.

\section{Orbiting hotspots}

Orbiting hot spots are compact, localized regions of enhanced radiation within the accretion disk of a black hole \cite{Narayan:2021, Gelles:2021}. These features emerge as a result of magnetic reconnection, plasma instabilities, or turbulence in the surrounding environment \cite{Narayan:2021}. In our formulation, the hot spot is treated as a point-like emitter following a circular, counterclockwise, equatorial geodesic orbit. This idealized model does not account for internal effects such as shearing and cooling, which may alter the emission characteristics of the hot spot.

Studying the motion and polarization of such hot spots provides valuable insights into the underlying spacetime geometry and potential deviations induced by dark matter effects with respect to the Schwarzschild geometry. In particular, our main motivation is to investigate how the dark matter component of the galactic halo influences the Stokes polarization parameters, modifying the observed polarization loops. These effects could serve as a potential probe for the large-scale distribution of dark matter around supermassive black holes, revealing its influence on the surrounding plasma and spacetime curvature on galactic scales.

\subsection{Polarization properties of the hot spot}
A hot spot in orbit moves along a circular geodesic within the equatorial plane, with its motion measured relative to a locally non-rotating frame (LNRF) in the presence of a dark matter halo. In a Schwarzschild spacetime, the innermost stable circular orbit (ISCO) is positioned at $r = 6M$. However, the addition of a dark matter halo modifies the ISCO radius, as described by Eq. (\ref{rISCO_GDM}), which in turn alters the orbital velocity profile. Specifically, for large compactness values of the halo, the ISCO radius decreases compared to its Schwarzschild counterpart, leading to stronger gravitational effects near the black hole.

Assuming a Keplerian motion of the hot spot around the central object, we consider circular orbits governed by the underlying spacetime geometry, (\ref{eq-lineelement}). The Keplerian orbital angular velocity at a radial coordinate $r=r_s$ is described by
\begin{equation}\label{Omega_K}
\Omega_K(r) = \frac{d\phi}{dt} = \left( \frac{f'(r)}{2r} \right)^{1/2},
\end{equation}
where the metric function $f(r)$ encodes the dependence on the total mass of the dark matter halo $M$ and its length scale $a_0$. The relativistic orbital velocity of the hot spot, as measured in the locally non-rotating frame (LNRF) along the azimuthal direction, can then be expressed as
\begin{equation}
\beta = \frac{r}{\sqrt{f(r)}}\,\Omega_K(r) = \left( \frac{r}{2} \log'\big(f(r)\big) \right)^{1/2}, \quad \chi = -\pi/2.
\label{v_LNRF_DM}
\end{equation}
The change in orbital velocity affects both the motion and the Doppler-shifted emission of the hot spot, which in turn results in observable changes in its polarization properties.

The analysis of the polarization of the hot spot is performed most effectively using the Stokes parameters $Q$ and $U$, defined according to \cite{Gelles:2021} as

\begin{equation}\label{QUParameters}
  Q=E_{obs}^{y}-E_{obs}^{x}, \quad U=-2E_{obs}^{x}E_{obs}^{y},
\end{equation}
where $E_{obs}^{x}$ and $E_{obs}^{y}$ are the observed electrical field components, (\ref{pol_EVPA}). It is important to emphasize that our pure synchrotron radiation model assumes a total polarization of unity, expressed as $\sqrt{Q^2+U^2}=I$.

The evolution of the hot spot motion in the $Q$--$U$ plane forms characteristic polarization loops, capturing the time-dependent structure of the screen polarization. These loops contain both the degree of polarization and the intensity of the polarized flux. The orientation of the electric vector position angle (EVPA) is directly linked to the $Q$--$U$ phase space and is determined by
\begin{equation}
    EVPA = \frac{1}{2} \arctan \left( \frac{U}{Q} \right).
\end{equation}
When a dark-matter halo is present, deviations in the metric modify the polarization loops in the $Q$--$U$ space, which can serve as indirect evidence of dark matter interactions. Stokes parameters $Q$ and $U$ are computed according to eq. (\ref{QUParameters}), assuming that the hot spot is an isotropic circular source rather than a thin disk, with redshift corrections included but excluding the contribution of geodesic path length $l_{p}$, introduced in equation (\ref{LenghtPath}). 

\begin{figure}[t!]
    \centering
    \includegraphics[width=\textwidth]{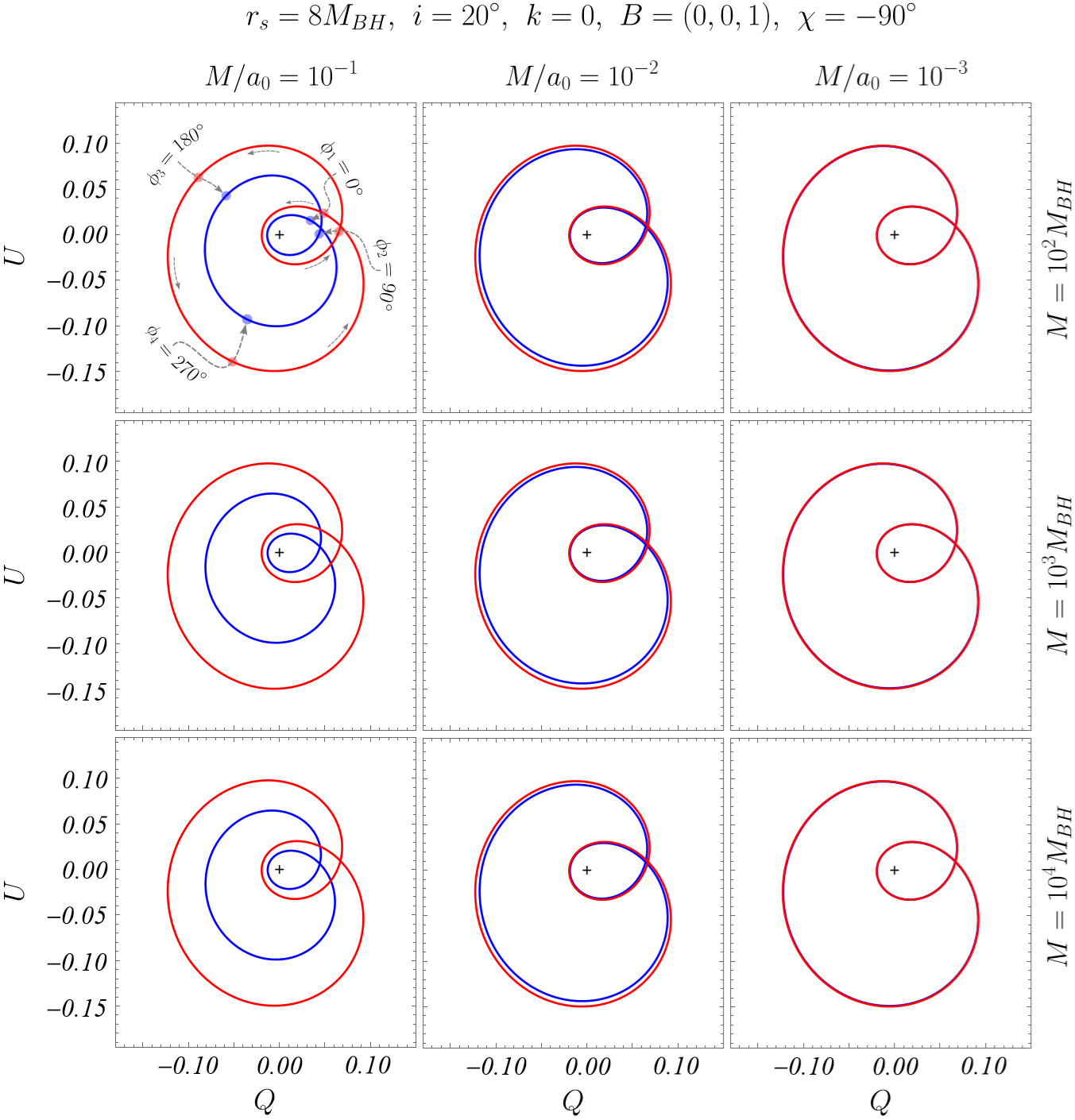}
    \caption{The plots illustrate the dependence of the polarized flux $Q$ versus $U$ for direct images under a vertical magnetic field and an inclination angle of $\theta_o = 20^\circ$. The results for Schwarzschild black holes and those with a dark matter halo are shown in red and blue, respectively. The rows from top to bottom correspond to three different dark matter halo masses: $M = 10^{2}M_{BH}$, $M = 10^{3}M_{BH}$, and $M = 10^{4}M_{BH}$. For each case, four values of halo compactness, $M/a_0$, namely $10^{-1}$, $10^{-2}$, and $10^{-3}$, are considered. The emission radius is fixed at $r_s = 8 M_{BH}$. In the top-left panel, the red and blue transparent dots indicate the azimuthal emission coordinates $\phi$, spaced by $90^\circ$, while the central black crosshairs mark the origin of each plot.}
    \label{fig:UQ_(rs=8.M)(k=0)(i=20deg)}
\end{figure}

\begin{figure}[t!]
    \centering
    \includegraphics[width=\textwidth]{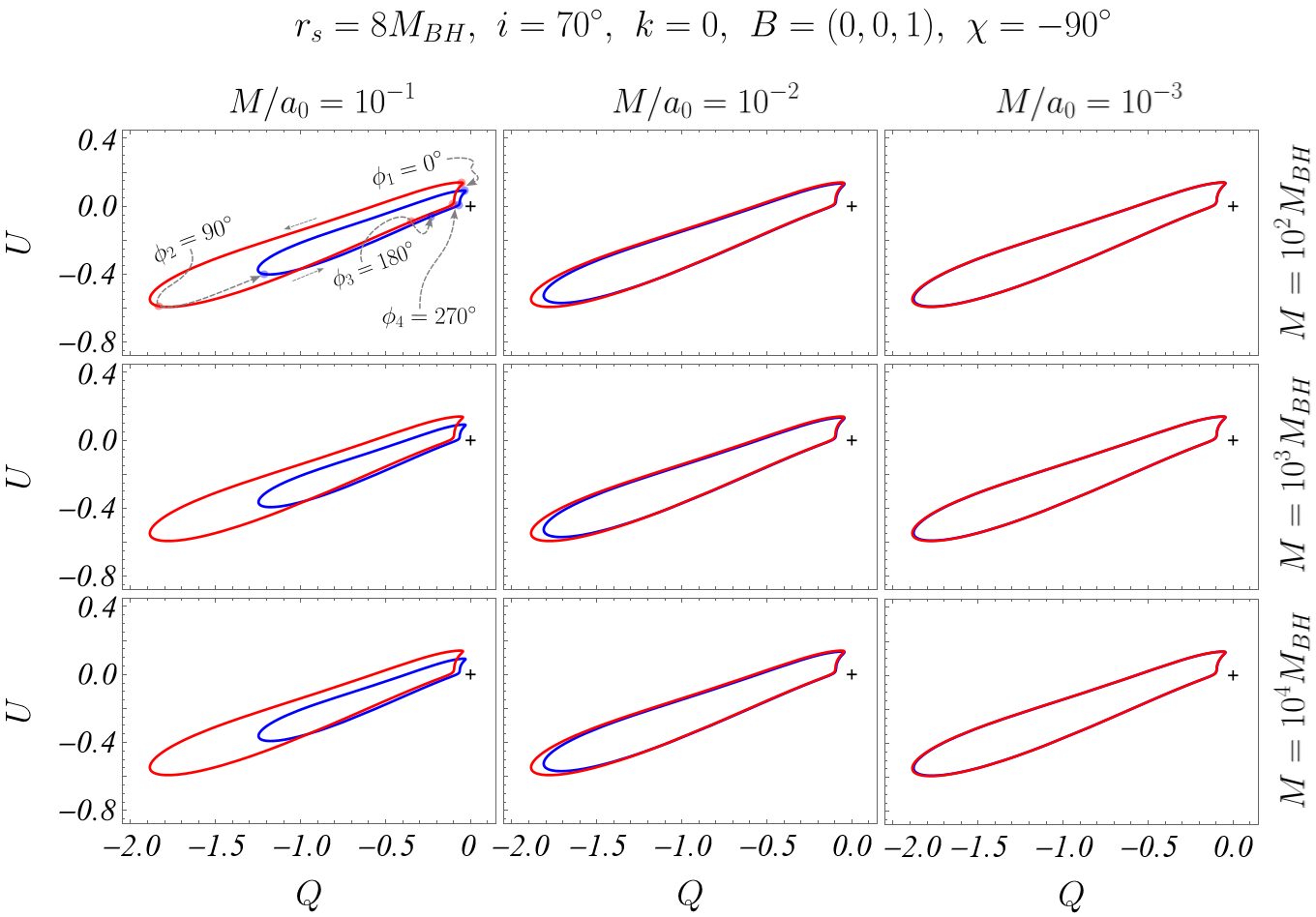}
    \caption{The plots illustrate the polarized flux $Q$ versus $U$ for direct images with a vertical magnetic field at an inclination of $\theta_{o} = 70^\circ$. Schwarzschild and dark matter halo black holes are shown in red and blue, respectively. Rows correspond to different halo masses ($M=10^{2}M_{BH}$ to $10^{4}M_{BH}$) and four compactness values ($M/a_{0} = 10^{-1}$ to $10^{-3}$). The emission radius is fixed at $r_{s} = 8 M_{BH}$. In the top left panel, black dots mark four azimuthal coordinates $\phi$ at $90^\circ$ intervals, with black crosshairs denoting the plot origins.}
    \label{fig:UQ_(rs=8.M)(k=0)(i=70deg)}
\end{figure}

\subsubsection{Polarization of direct images}

Figures \ref{fig:UQ_(rs=8.M)(k=0)(i=20deg)} and \ref{fig:UQ_(rs=8.M)(k=0)(i=70deg)} show loops in the $Q$--$U$ plane for inclination angles of $i = 20^\circ$ and $70^\circ$, respectively. In each case, the contours are obtained by numerically integrating the Hamilton equations and extracting the observed polarized intensity of the direct emission from a hot spot, according to equations (\ref{pol_EVPA}), under the assumption of a completely vertical magnetic field.

Each figure contains panels for two cases: a Schwarzschild black hole and a black hole surrounded by a dark matter halo, depicted in red and blue, respectively. The configurations further explore three values of the dark-matter halo compactness $M/a_0$, namely $10^{-1}$, $10^{-2}$, and $10^{-3}$, arranged from left to right. These are considered for halo masses $M/M_{\rm BH} = 10^2$, $10^3$, and $10^4$, ordered from top to bottom.

The orbit of the emitting hot spot has a radius of $r_s = 8M_{\rm BH}$, which lies outside the innermost stable circular orbit (ISCO) for both spacetimes. Additionally, the upper-left panel of each figure marks four azimuthal emission angles $\phi$, spaced $90^\circ$ apart along the orbit for both metrics. These angles are calculated based on the fact that geodesics in the spacetime of non-rotating black holes lie in a plane, assuming an initial condition $\phi = 0$, aligned with the direction toward the observer. 

In orbital motion around the Schwarzschild black hole, the relativistic orbital velocity of the hot spot at the chosen orbital radius is $\beta \simeq 0.4082$. In a geometry that includes the presence of dark matter, the velocity remains nearly unchanged, even at high compactness of the galactic halo, but increases slightly to $\beta \simeq 0.4092$ for $M/a_{0} = 10^{-1}$ with a halo mass of $M = 10^2\,M_{\rm BH}$. On the other hand, as the compactness $M/a_0$ of the dark matter halo decreases, the relativistic orbital velocity asymptotically approaches the Schwarzschild value, as expected. Specifically, the velocity decreases for higher halo masses and increases slightly for lower halo masses.

Although the relativistic velocity increases slightly at high compactness values of the galactic dark matter halo, no noticeable change is observed in the topology of the $Q$--$U$ contours. On the contrary, their shape remains closely aligned with that of the Schwarzschild black hole.

For a given hot spot orbit, increasing the compactness of the dark matter halo consistently results in lower polarization intensity. The higher the compactness, the more compact the $Q$--$U$ contours become, compared to those produced in the Schwarzschild geometry.

As expected, in the limit of vanishing compactness in the dark matter distribution ($M/a_0 \rightarrow 0$), the polarization level increases, gradually approaching the asymptotic values marked by the red contours. Across all configurations, variations in halo mass $M$ lead to only subtle changes in the shape and size of the contours, as clearly illustrated in the Figures.

\subsubsection{Polarization of indirect images}

\begin{figure}[t!]
    \centering
    \includegraphics[width=\textwidth]{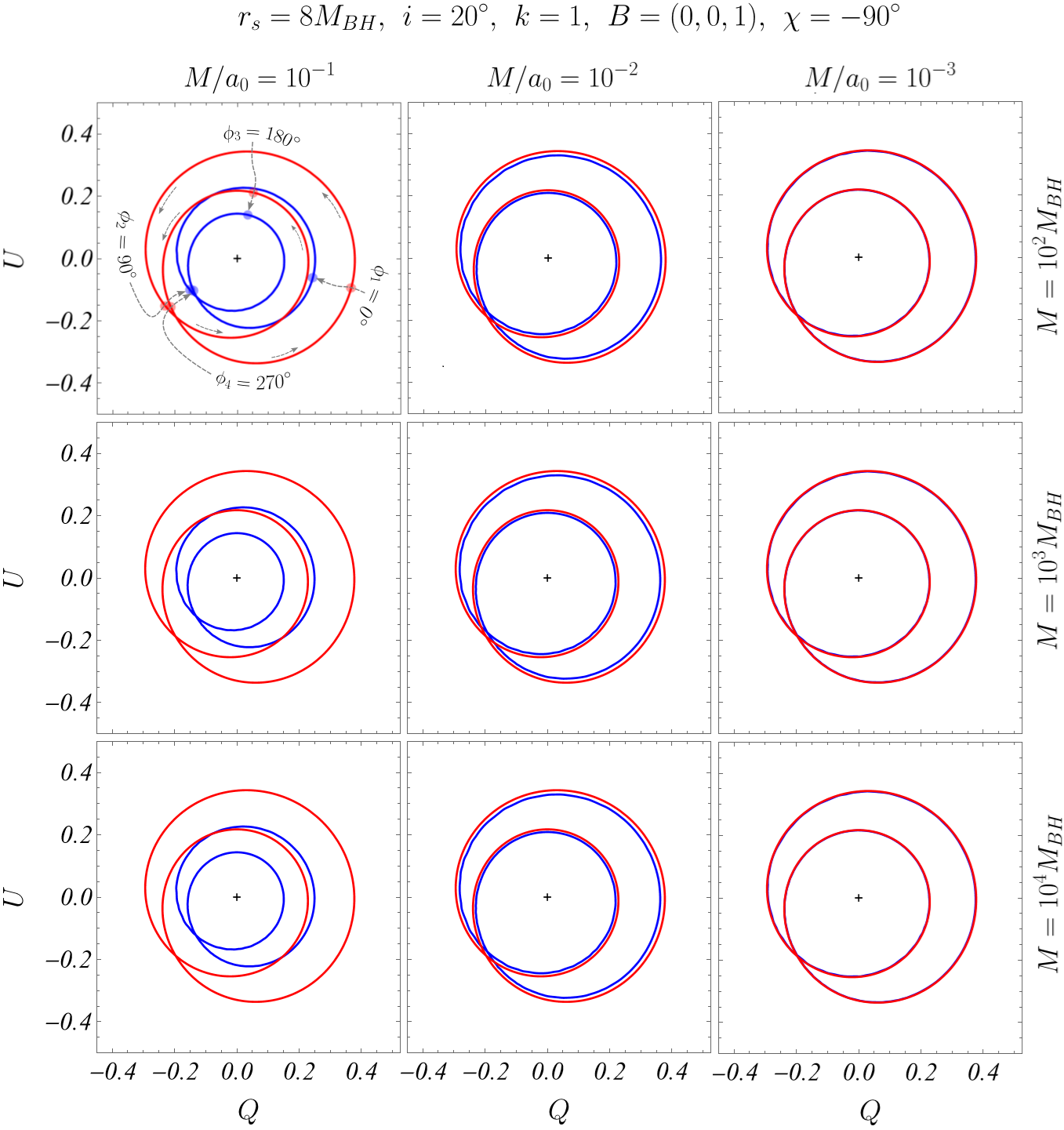}
    \caption{The plots depict the relationship between the polarized flux $Q$ and $U$ for indirect images of order $k=1$ under a vertical magnetic field with an inclination angle of $\theta_{o} = 20^\circ$. Schwarzschild black holes and those with a dark matter halo are represented in red and blue, respectively. The rows correspond to different halo masses, ranging from $M=10^{2}M_{BH}$ to $10^{4}M_{BH}$, and four compactness values ($M/a_{0} = 10^{-1}$ to $10^{-3}$). The emission radius is set at $r_{s} = 8 M_{BH}$. In the top left panel, black dots indicate four azimuthal emission coordinates $\phi$ at $90^\circ$ intervals, while black crosshairs mark the plot origins.}
    \label{fig:UQ_(rs=8M)(k=1)(i=20deg)}
\end{figure}

\begin{figure}[t!]
    \centering
    \includegraphics[width=\textwidth]{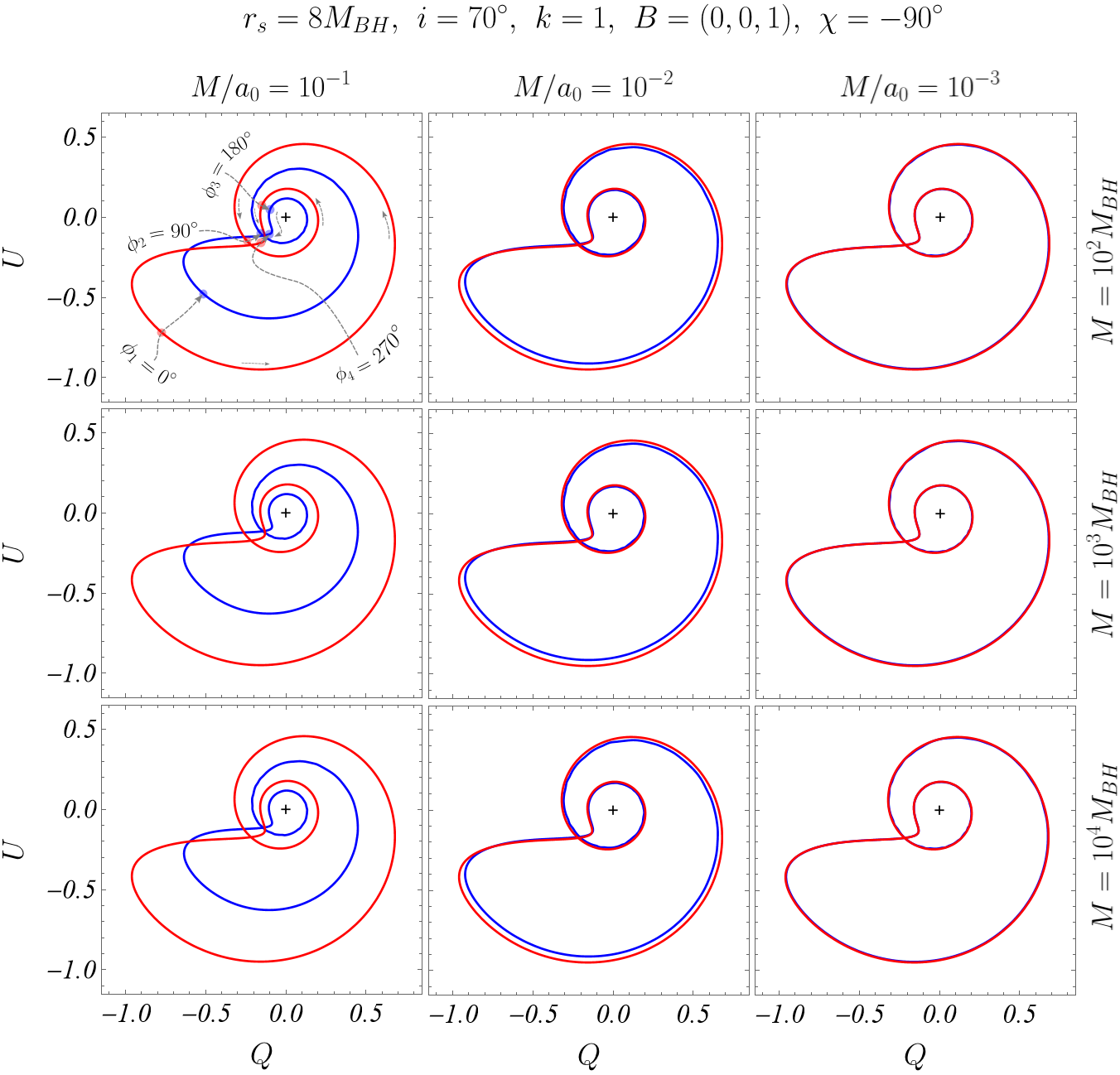}
    \caption{The plots illustrate the dependence of the polarized flux $Q$ versus $U$ for indirect images of order $k=1$, considering a vertical magnetic field and an inclination angle of $\theta_{o} = 70^\circ$. Schwarzschild black holes and those surrounded by a dark matter halo are shown in red and blue, respectively. The rows represent different halo masses, varying from $M=10^{2}M_{BH}$ to $10^{4}M_{BH}$, along with four compactness values ($M/a_{0} = 10^{-1}$ to $10^{-3}$). The emission radius is fixed at $r_{s} = 8 M_{BH}$. In the top left panel, black dots mark four azimuthal emission coordinates $\phi$ at $90^\circ$ intervals, while black crosshairs indicate the plot origins.}
    \label{fig:UQ_(rs=8M)(k=1)(i=70deg)}
\end{figure}

The polarization $Q$--$U$ diagrams in indirect images of order $k = 1$ exhibit the same qualitative behavior across different compactness values and halo masses of the galactic dark matter component. At relatively high compactness, dark matter significantly suppresses the polarization fraction, and the $Q$--$U$ loops remain entirely enclosed within the region defined by the Schwarzschild case. As the dark matter content decreases, the polarization increases, approaching the level characteristic of a Schwarzschild black hole. In this regime, increasing the mass of the dark matter halo has only a minimal effect, as illustrated in Figures \ref{fig:UQ_(rs=8M)(k=1)(i=20deg)} and \ref{fig:UQ_(rs=8M)(k=1)(i=70deg)}.

One particularly interesting feature is that the $Q$--$U$ loops in the indirect images always remain strongly extended along both axes, regardless of the observer's inclination. However, at higher inclinations, the polarization exhibits a greater spread along the $Q$ direction compared to the $U$ direction, as seen in Figure \ref{fig:UQ_(rs=8M)(k=1)(i=70deg)}. A characteristic property of the $Q$--$U$ loops at $k = 1$ is that they consistently encircle the origin. 

\subsection{$Q, U$ Loop Topology}

The topology of the $Q$--$U$ loops depends on the image type within the spacetime geometry of the galactic dark matter halo. We begin by examining the loop structure in direct images of the hot spot.

\subsubsection{Topology of $Q,U$ loops in direct images}

In addition to the effects of compactness and halo mass, the topology of the polarization loops in the $Q$--$U$ plane is also sensitive to the observer’s inclination. For a relatively low inclination of $i = 20^\circ$, the diagrams consistently display two distinct loops traced by the orbiting hot spot, regardless of the compactness parameter.

At a higher inclination angle, specifically at $i = 70^\circ$, the $Q$--$U$ contours display a different topological structure, as shown in Figure \ref{fig:UQ_(rs=8.M)(k=0)(i=70deg)}. The number of loops reduces to one, and it adopts a strongly elongated shape in the $Q$-direction, as first reported in \cite{Gelles:2021}. This elongation reflects enhanced polarization variability and highlights a shift in the electric vector position angle ($EVPA$), particularly in regions where significant light bending occurs--such as when photons are emitted from behind the compact object. Notably, a characteristic property of the $Q$--$U$ loops for the direct images is that they do not encircle the origin.

In this configuration, part of the photon flux originates to the right of the compact object, at azimuthal angles in the range $0^\circ\leq\phi\leq 90^\circ$, while the rest is emitted from the left side of the black hole when the azimuthal angle falls within the range $90^\circ\leq\phi\leq180^\circ$, as demonstrated in the upper-left panel of Figure \ref{fig:UQ_(rs=8.M)(k=0)(i=70deg)}. The most strongly polarized intensity is associated with photons emitted at azimuthal positions in the range $180^\circ\leq\phi\leq360^\circ$, where the gravitational redshift is at its maximum.

\subsubsection{Topology of $Q,U$ loops in indirect images}

The topology of the $Q$--$U$ loops in indirect images of the hot spot is largely independent of the observer’s inclination. At $i = 20^\circ$, two distinct loops appear regardless of the compactness value, and this two-loop structure persists even at $i = 70^\circ$. In indirect images, the loops not only continue to encircle the origin, but also exhibit a more pronounced deformation of the second loop at higher inclinations, unlike the $Q$--$U$ loops observed at lower angles. Notably, the polarization intensity in these indirect images is approximately twice as low as that in the corresponding direct images.

The strongest polarization is associated with photons emitted at azimuthal positions around $\phi \sim 0^\circ$ or $360^\circ$, where the gravitational redshift reaches its maximum. When the hot spot moves through the interval $-90^\circ \leq \phi \leq 90^\circ$, it generates the first loop in the $Q$--$U$ plane, while the second loop forms during the remaining part of the orbit, $90^\circ \leq \phi \leq 270^\circ$.

\section{Conclusion}

Uncovering the nature of dark matter is an important challenge in fundamental physics. While constraints on its properties are mainly expected from cosmological experiments, electromagnetic observations in the strong field regime may offer further insights on its gravitational interaction. On the other hand, the back reaction of dark matter on the gravitational field of compact objects will have impact on the electromagnetic observables. This impact is expected to be small but with the increasing resolution of the electromagnetic missions it may not be negligible.

In this paper we study the influence of the dark matter halo in galaxies on the observable polarization from the accretion disk around the supermassive compact object in their center. The gravitational interaction of the dark matter halo with the central compact object is described by considering a spherically symmetric solution of the Einstein equations with an anisotropic matter distribution. At small scales the spacetime describes a generalization of the Schwarzschild black hole while at large distances corresponds to a Hernquist-type density profile which is well-known to recover the galactic rotation curves.

Since we are interested in the emission from the innermost regions of the accretion disk we adopt a simplified toy model describing a thin fluid ring orbiting in the equatorial plane. The fluid is located in a constant magnetic field and emits synchrotron radiation. Despite its simplicity the model reproduces the observed linear polarization from M87* and the polarization properties of the hot spots around Sgr A*. We calculate the intensity and the EVPA of the observable polarization around the fluid ring and evaluate the impact of the dark matter halo on these quantities. For the purpose we consider different inclination angles and magnetic field configurations as well as direct and indirect lensed images. The dark matter distribution is characterized by its compactness parameter which is chosen within the typical values for galactic dark matter halos.

We obtain that in the direct images the polarization intensity and direction deviate with less than $1\%$ from the isolated Schwarzschild black hole. When we consider indirect images the deviation increase with an order of magnitude but still reaches less than $10\%$ for both polarization observables. This implies that it will be difficult to estimate the impact of the dark matter halo on the central black hole through observations of the polarized emission of the accretion disk.

The influence of the dark matter halo on the polarization properties of the direct emission may be comparable in magnitude to the effects produced by modification of the spacetime geometry by considering for example horizonless compact objects. Thus, the two sources of deviation from the predictions for the isolated Schwarzschild black hole may be difficult to disentangle when the necessary resolution is achieved. However, in the indirect images  we may observe with two orders of magnitude larger deviation from Schwarzschild for horizonless compact objects than due to the influence of the dark matter halo \cite{Nedkova:2023}, \cite{Delijski}. This suggests that the indirect images provide much finer tests for beyond Kerr phenomenology.

\section{Appendix}

Let us briefly examine how the polarization contours in the $Q-U$ plane change with variations in the orbital radius of the hotspot. As the orbital radius decreases from $r_s = 8M$ to $6M$ and $4M$, the relativistic velocity of the hotspot orbiting a black hole surrounded by a dark matter halo with compactness $M/a_0 = 10^{-1}$ and halo mass $M = 10^2 M_{\rm BH}$ increases slightly, taking values of approximately $\beta = 0.4092$, $0.5006$, and $0.7074$. These values are marginally higher than the corresponding ones for the Schwarzschild black hole: $\beta = 0.4082$, $0.5000$, and $0.7071$.

Despite the increase in relativistic velocity with decreasing orbital radius, at an inclination angle of $i = 20^\circ$, the magnitude of the observed polarized flux decreases. The $Q$--$U$ diagrams become increasingly dephased in the counterclockwise direction, as illustrated in Figure \ref{fig:UQ_(rs=4, 6, 8M)(k=0)(i=20.0deg)}. Notably, the polarization diagrams retain a two-loop structure at all considered orbital radii.

In contrast, at higher inclination angles, such as $i = 70^\circ$, the $Q-U$ diagrams become strongly elongated along the $Q$ direction, as depicted in Figure \ref{fig:UQ_(rs=4, 6, 8M)(k=0)(i=70.0deg)}. A distinct feature emerges: for outer orbits with relatively large radii ($r_s = 6M$ and $8M$), the diagrams exhibit only a single loop, whereas for the inner orbit ($r_s = 4M$), the polarization dynamics change significantly. In this case, the polarized flux varies more strongly in the $Q$ direction, and the $Q-U$ diagram develops a secondary, miniature loop that is not observed in the outer orbits at the same inclination. 

\begin{figure}[t!]
    \centering
    \includegraphics[width=\textwidth]{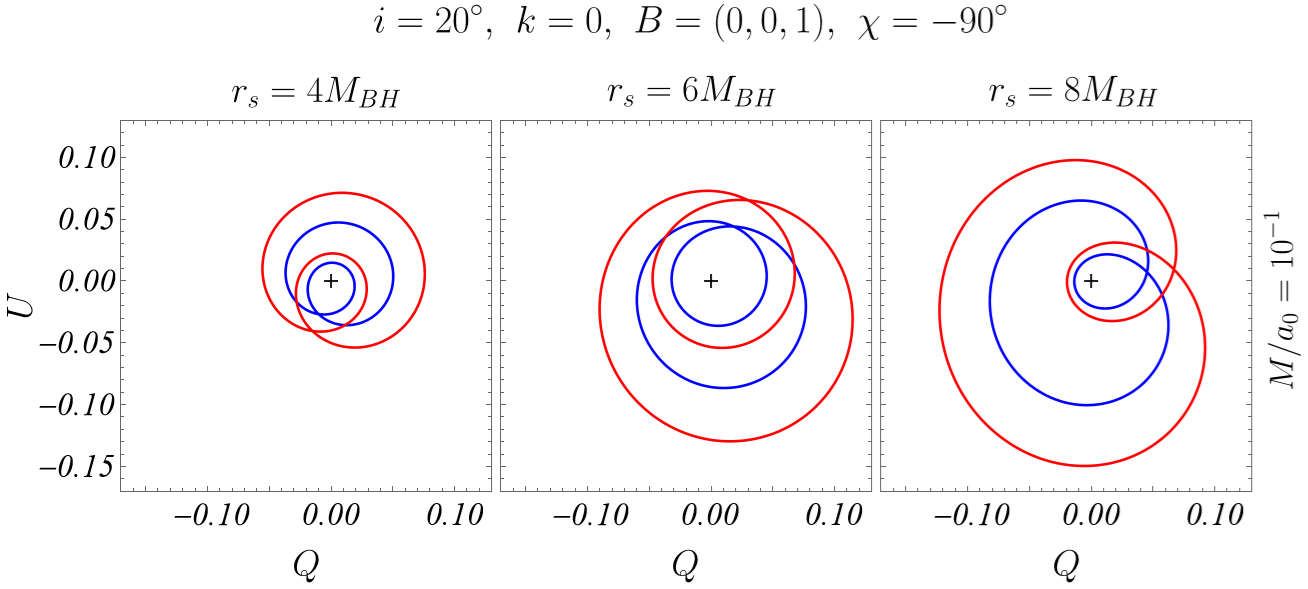}
    \caption{The plots illustrate the dependence of the polarized flux $Q$ versus $U$ for the direct hot spot image under a vertical magnetic field and an inclination angle of $\theta_o = 20^\circ$. The left, central, and right panels correspond to emission radii of $r_s = 4M_{BH}, 6M_{BH}$, and $8M_{BH}$, respectively. The results for Schwarzschild black holes and those with a dark matter halo are shown in red and blue, respectively. The dark matter halo has a mass of $M = 10^{2}M_{BH}$ and a compactness parameter of $M/a_0 = 10^{-1}$.}
    \label{fig:UQ_(rs=4, 6, 8M)(k=0)(i=20.0deg)}
\end{figure}

\begin{figure}[t!]
    \centering
    \includegraphics[width=\textwidth]{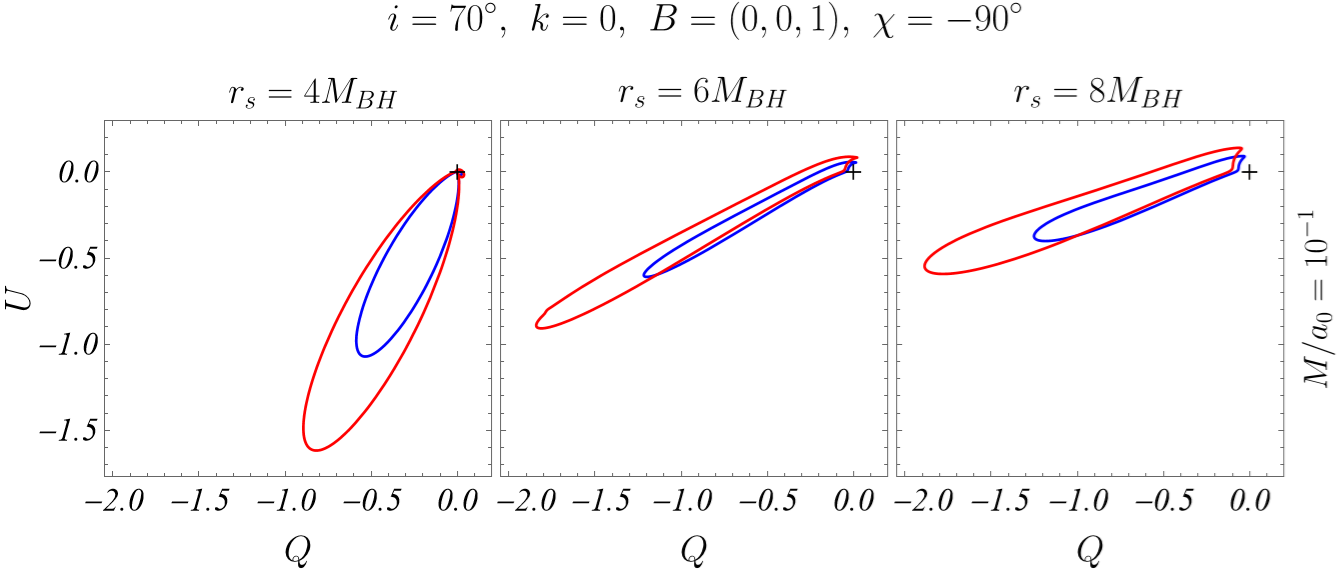}
    \caption{The graphs depict the relationship between the polarized flux $Q$ and $U$ for the direct hot spot image in the presence of a vertical magnetic field and at an inclination angle of $\theta_o = 70^\circ$. The left, middle, and right panels represent emission radii of $r_s = 4M_{BH}, 6M_{BH}$, and $8M_{BH}$, respectively. The results are displayed in red for Schwarzschild black holes and in blue for those surrounded by a dark matter halo. The dark matter halo is characterized by a mass of $M = 10^{2}M_{BH}$ and a compactness parameter of $M/a_0 = 10^{-1}$.}
    \label{fig:UQ_(rs=4, 6, 8M)(k=0)(i=70.0deg)}
\end{figure}

Moreover, in indirect images of order $k = 1$, the polarization patterns reveal an intriguing trend. For outer orbits at $r = 8M$ and relatively small inclination angles, the polarization flux in the Schwarzschild geometry exceeds that of the corresponding direct images by a factor of approximately 2.52. The geometric model of a black hole with a dark matter halo follows a similar trend, where the polarization intensity in indirect images is about 2.48 times greater than that in the direct images.

\begin{figure}[t!]
    \centering
    \includegraphics[width=\textwidth]{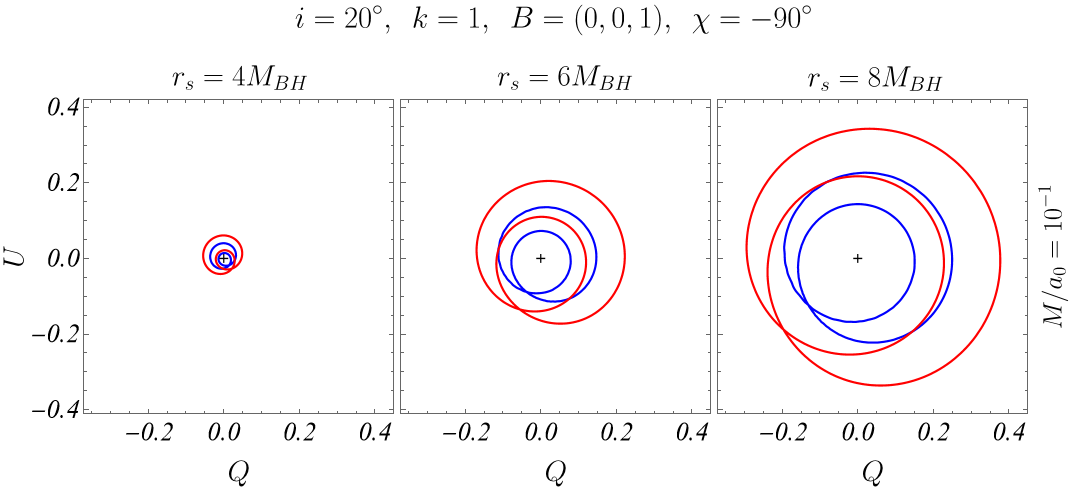}
    \caption{The graphs illustrate the relationship between the polarized flux $Q$ and $U$ for the direct hot spot image in the presence of a vertical magnetic field and at an inclination angle of $\theta_o = 20^\circ$. The left, middle, and right panels correspond to emission radii of $r_s = 4M_{BH}, 6M_{BH}$, and $8M_{BH}$, respectively. The results are shown in red for Schwarzschild black holes and in blue for those surrounded by a dark matter halo. The dark matter halo is characterized by a mass of $M = 10^{2}M_{BH}$ and a compactness parameter of $M/a_0 = 10^{-1}$.}
    \label{fig:UQ_(rs=4, 6, 8M)(k=1)(i=20.0deg)}
\end{figure}

\begin{figure}[t!]
    \centering
    \includegraphics[width=\textwidth]{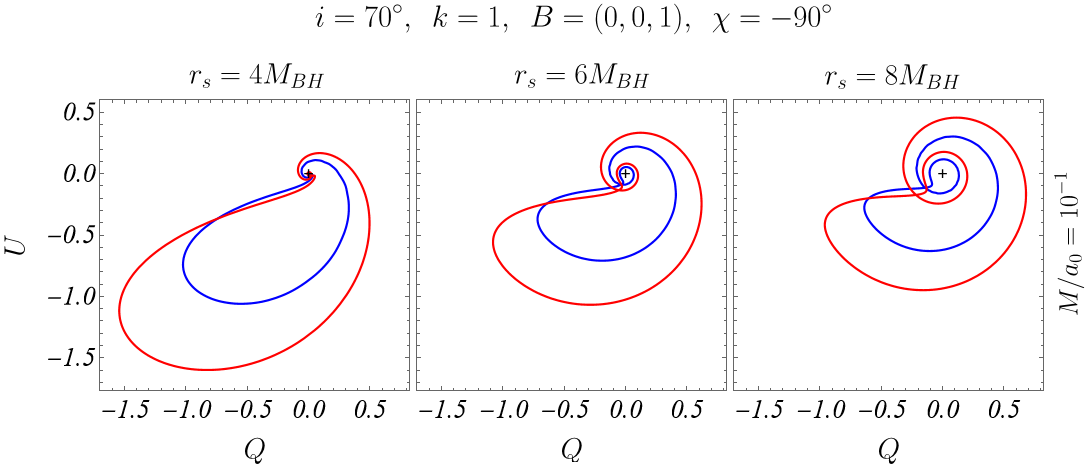}
    \caption{The graphs represent the correlation between the polarized flux $Q$ and $U$ for the direct hot spot image in the presence of a vertical magnetic field at an inclination angle of $\theta_o = 70^\circ$. The left, middle, and right panels depict emission radii of $r_s = 4M_{BH}, 6M_{BH}$, and $8M_{BH}$, respectively. The results are illustrated in red for Schwarzschild black holes and in blue for those encompassed by a dark matter halo. The dark matter halo is defined by a mass of $M = 10^{2}M_{BH}$ and a compactness parameter of $M/a_0 = 10^{-1}$.}
    \label{fig:UQ_(rs=4, 6, 8M)(k=1)(i=70.0deg)}
\end{figure}

On the other hand, for a halo with compactness $M/a_0 = 10^{-1}$ and mass $M = 10^2 M_{\rm BH}$, the polarization flux is approximately 1.51 times lower than that of the Schwarzschild case, as shown in Figures \ref{fig:UQ_(rs=4, 6, 8M)(k=0)(i=20.0deg)} and \ref{fig:UQ_(rs=4, 6, 8M)(k=1)(i=20.0deg)}.

Furthermore, decreasing the orbital radius of the hot spot from $r_{s} = 8M$ through $6M$ down to $4M$ leads to a significant reduction in polarization intensity in both geometries (with and without dark matter). For indirect images, the intensity becomes about $0.79$ times lower than that in direct images. This result highlights a key transition: for orbits sufficiently close to the event horizon, the direct images become more strongly polarized than the indirect ones. This is in contrast to the behavior at larger orbital radii, where the indirect images generally exhibit stronger polarization.

Importantly, the presence of dark matter does not alter this trend but progressively suppresses the overall polarization intensity as the halo becomes more compact. In all examined cases across different orbital radii, the topology of the $Q-U$ contours remains unchanged: the contours are consistently dephased in a clockwise direction and retain a two-loop structure.

Moreover, with an increase in the observer's inclination to $i = 70^\circ$, no unexpected behavior is observed in the polarization flux. As the orbital radius of the hot spot decreases, the $Q$--$U$ diagrams become increasingly elongated, and the maximum polarization intensity grows monotonically, as clearly shown in Figure \ref{fig:UQ_(rs=4, 6, 8M)(k=1)(i=70.0deg)}.

At the same time, for all studied orbits, the topology of the $Q$--$U$ contours remains characterized by two loops. However, as the orbital radius decreases, the second loop becomes progressively more contracted, while the first loop expands significantly. This expansion corresponds to strongly polarized photons emitted by the hot spot when it is located behind the compact object during its orbital motion.

\section*{Acknowledgments}
We gratefully acknowledge support by the Bulgarian NSF Grant KP-06-H68/7.

\end{document}